\documentclass[usenatbib]{mn2e}
\usepackage{amssymb}
\usepackage{graphicx}
\title[GW from cosmological binaries]{The Gravitational Wave Background from Cosmological Compact Binaries}

\begin{document}
\author[A.J.~Farmer and E.S.~Phinney]{Alison J.~Farmer\thanks{e-mail:
    ajf@tapir.caltech.edu} and E. S.~Phinney\\
Theoretical Astrophysics, MC 130-33 Caltech, Pasadena, CA 91125, USA}
\maketitle

\begin{abstract}
We use a population synthesis approach to characterise, as a function of
cosmic time, the extragalactic close binary population descended from
stars of low to intermediate initial mass. The unresolved gravitational wave
(GW) background due to these systems is calculated for the 0.1--10 mHz
frequency band of the planned Laser Interferometer Space Antenna
(LISA). This background is found to be dominated by emission from
close white dwarf--white dwarf pairs. The spectral shape can be understood in terms of
some simple analytic arguments. To quantify the astrophysical
uncertainties, we construct a range of
evolutionary models which produce populations consistent with Galactic
observations of close WD--WD binaries. The models differ in binary
evolution prescriptions as well as initial parameter distributions and
cosmic star formation histories. We compare the resulting
background spectra, whose shapes are found to be insensitive to the
model chosen, and different to those found recently by Schneider et al. (2001). From this set of models, we constrain the amplitude of
the extragalactic background to be $1 \times 10^{-12} \la \Omega_{\rm{gw}}
\rm{(1~mHz)} \la 6 \times 10^{-12}$, in terms of $\Omega_{\rm{gw}}(f)$, the
fraction of closure density received in gravitational waves in the
logarithmic frequency interval around $f$.

\end{abstract}


\section{Introduction}

Except at very low radio frequencies, most electromagnetic telescopes
have good angular rejection, so that faint sources and backgrounds can
be seen by looking between bright sources.
In contrast all currently implemented gravitational wave detectors, and most 
of those envisaged for the future, simultaneously respond to sources all
over the sky, modified only by a beam pattern of typically quadrupole form.
It is therefore important to understand the brightness of the gravitational
wave sky, since this will limit the ultimate sensitivity attainable
in gravitational wave astronomy.
One immediate pressure to understand this background comes from the
need to set design requirements for the ESA/NASA 
Laser Interferometer Space Antenna (LISA) mission
(LISA mission documents and status may be found at
http://lisa.jpl.nasa.gov/ and http://sci.esa.int/home/lisa/).

In this paper, we attempt to predict the gravitational wave background
produced by all the binary stars in the universe, excluding neutron
stars and black holes.  This is believed to be the principal source of
gravitational wave background in the frequency range
$10^{-5}<f<10^{-1}$ Hz.  Below $10^{-5}$ Hz, the background is
probably dominated by merging supermassive black holes, and above
$10^{-1}$ Hz, it is probably dominated by merging neutron stars and
stellar mass black holes (whose complicated and poorly understood
formation histories and birth velocities make predictions more
uncertain, cf. \citealt*{bel02}).  

Besides the extragalactic background,
there is also a Galactic background produced by the binary stars in our
Milky Way (\citealt*{eva87}; \citealt*{hil90}; \citealt*{nel01c}).
Although the Galactic background is many
times larger in amplitude
than both the extragalactic background and LISA's design sensitivity, 
the individual binaries contributing to it can
be (spectrally) resolved and removed at frequencies above 
$\sim 3\times 10^{-3}$ Hz \citep{cor02}.  Below this frequency
they cannot be removed (at least in a mission of reasonable lifetime
$\sim 3$ years), but the unresolved Galactic background will be
quite anisotropic.  As the detector beam pattern rotates about the sky,
the Galactic background will thus be modulated, while the isotropic
(or nearly so; see \citealt{kos00}) distant extragalactic background
will not.  Modelling of the angular distribution of the
Galactic background using both {\em a priori} models and the observed 
distribution of higher frequency resolved sources
will thus allow the Galactic background to be subtracted to some
precision (\citealt{gia97}). In addition, there will be anisotropies
due to the distribution of local galaxies, at the level of 10 per cent
of the distant extragalactic background from the LMC, and at the
per cent level from M31 or the Virgo cluster (see also \citealt{lip95}).

The immediate motivation for this work is a design issue
for LISA.
One of LISA's major science goals (see the LISA Science Requirements 
document
at http://www.tapir.caltech.edu/listwg1/) is the detection
of gravitational waves from compact objects spiralling into
supermassive black holes (\citealt{fin00}; \citealt{hil95}), since these can provide precision tests of strong field
relativity and the no-hair theorem \citep{hug01}.  However, these
signals are weak, and their templates not yet fully understood. It has
thus been proposed that LISA should be designed with somewhat greater
sensitivity to increase the probability that these signals are
detected.  However, this would be pointless if the principal
background were cosmological rather than instrumental.  As we shall
see (Fig. \ref{fig:lisa}), we find that this is most probably almost, but not
quite the case at the relevant frequencies ($4-10$ mHz). So there
would be a point to increasing LISA's sensitivity in the $4-10$ mHz
range, but not to increasing it by more than a factor of 3 in
gravitational wave amplitude $h$ (9 in $\Omega\propto f^2 h^2$).

A second motivation for this work comes from the fact that this
background is an astrophysical {\em foreground} to searches (both with
LISA and with future detectors with extended frequency range and
sensitivity) for backgrounds produced in the very early universe.
Gravitational waves from bubble walls and turbulence following the
electroweak phase transition are expected to be in the LISA frequency
band, with amplitude that could be well above LISA instrumental
sensitivity (\citealt*{kam94}; \citealt*{kos02}; \citealt{apr02}). Another potential source of
isotropic gravitational waves in the LISA band are those produced when
dimensions beyond the familiar four compactified, which occurred when the
universe had temperature $kT>$TeV \citep{hog00}.

Note that detection of a gravitational wave background can possibly be
made even if it is considerably below the noise limit of the LISA
detectors shown in our Fig. \ref{fig:lisa}.  This can be done by comparing the
signals from Michelson beam combinations (sensitive to instrument
noise and gravitational waves) with Sagnac beam combinations
(sensitive to instrument noise, but insensitive to gravitational
waves), thus calibrating the instrumental noise ---cf. \citet*{tin01},
\citet{hog01}.

Gravitational waves are the only directly detectable relic
of inflation in the early universe, and their detection over a range
of frequencies would provide a valuable test of models of inflation
\citep{tur97}.  It has been proposed that advanced space-based
gravitational wave detectors might search for the background of
gravitational waves from inflation.
The gravitational waves from slow-roll inflation models
contribute to the critical density in the universe
$\Omega_{\rm{gw}} < 10^{-15}$ per octave of frequency.
We shall see (Fig. \ref{fig:wdtypes}, \ref{fig:nsources}) 
that the gravitational wave background from cosmological
binaries makes such detection impractical except
at frequencies below $10^{-5}$ Hz (where supermassive black holes
continue to make it impossible), or above $0.1$ Hz.

A third motivation is that a detection of the extragalactic
binary background, e.g. by
LISA, would set an independent (and unaffected by dust extinction)
constraint on a combination of the star formation history of the
universe and binary star evolution.

There have been previous estimates of the extragalactic binary
background.  \citet*{hil90} made detailed estimates
of the Galactic binary background, and estimated that the
extragalactic background from close double white dwarf pairs should be about 2 per cent (in flux or $\Omega$
units) of the Galactic background.  This estimate was refined, using
more modern star formation histories, by \citet{kos98}, who found
instead a level of $\sim$10 per cent. \citet{sch01} used a descendant of the Utrecht
population synthesis code to estimate the extragalactic binary
background as a function of frequency, and claimed that the background
should have a large peak at $\sim 3\times 10^{-5}$ Hz, just below the frequency
at which typical binaries have a lifetime that equals the age of the
Universe.

We have followed the spirit of this previous work, but with an
independent binary population synthesis code. More importantly,
we have devoted much effort to the normalisation of the background,
to understanding the contributions of different types of binaries
and their formation pathways to the background, and to estimating
the uncertainties in all of these, so that we can have a
better idea of the sources and level of uncertainty in the
predicted background.  

The paper is organised as follows: In section \ref{sec:lgw}, we describe the
gravitational wave (GW) emission from a binary system, then in section
\ref{sec:evol} we outline the main evolutionary pathways to the close double
degenerate (DD) stage, which we shall see is the dominant source of
GW background in the LISA band. In section \ref{sec:anal}, we use the preceding sections to make
some simple analytic arguments about the nature of DD inspiral
spectra. We describe the use of the \textsc{bse} code in our population
synthesis, in section \ref{sec:test}, then go on to construct a set of synthesis
models whose results we test against the observed Galactic DD
population. We also motivate some modifications made to the
prescription for the evolution of AM CVn stars in the \textsc{bse}
code. In section \ref{sec:cosmocalc}, we present the cosmological integrals used in the
code, along with the cosmic star formation history and overall normalisation chosen. Section \ref{sec:results} is
devoted to a discussion of the GW background spectra produced by our
code, in terms of the systems contributing to the background and the
progenitors of these sources. We also discuss the differences between
our population synthesis models. In section \ref{sec:opt_pes}, we place limits on the
maximum and minimum expected background signals, and compare these
with the LISA sensitivity and in section \ref{sec:prev} with previous work. In section \ref{sec:conclusion} we summarise and conclude.

\section{Gravitational waves from a binary system}
\label{sec:lgw}

A binary system of stars in circular orbit with masses $M_1$ and $M_2$ and
orbital separation $a$ emits gravitational radiation, at the expense
of its orbital energy, at a rate given by \citep{pm63}
\begin{eqnarray}
L_{\rm{circ}}&=&\frac{32}{5}\frac{G^4}{c^5}\frac{(M_1
  M_2)^2(M_1+M_2)}{a^5} \nonumber\\
&\simeq& 1.0 \times 10^{32} \frac{ (M_1'
  M_2')^2(M_1'+M_2')}{(a')^5} \textrm{ erg s}^{-1},\label{eq:lcirc}
\end{eqnarray}
where primes denote quantities expressed in solar units,
  i.e. $M/\rm{M}_\odot$, $a/\rm{R}_\odot$. The gravitational
  radiation is emitted at twice the orbital frequency $\nu$ of the binary,
  $f_{\rm{circ}} = 2 \nu = \Omega/\pi$.

If the binary is eccentric with eccentricity $e$, this expression
must be generalised to include emission at all harmonics $n$ of the orbital frequency,
$f_n=n\nu = n\Omega/2\pi$, where $\Omega=(a^{-3} G(M_1+M_2))^{1/2}$. The luminosity in each harmonic is given
by
\begin{equation} L(n,e)=g(n,e)L_{\rm{circ}}, \label{eq:harm} \end{equation}
where $L_{\rm{circ}}$ is the luminosity of a circular binary with
separation $a$, as given in Eq. \ref{eq:lcirc}, where $a$ is now the relative
semi-major axis of the eccentric orbit, and the
$g(n,e)$ are defined in eq. (20) of \citet{pm63}. The total
specific luminosity  $L_{f} = dL(f)/df$ of the system is then a
sum over all harmonics:
\begin{equation}L_{f}(e)= L_{\rm{circ}} \sum_{n=1}^{\infty} g(n,e)
  \delta(f-n\nu). \label{eq:sumharm}\end{equation} The total luminosity is
\begin{equation}
L= L_{\rm{circ}} \sum_{n=1}^\infty g(n,e) = \frac{
  1+\frac{73}{24}e^2+\frac{37}{96}e^4}{(1-e^2)^{7/2}} L_{\rm{circ}} \label{eq:lgw}\end{equation}

For eccentric orbits, the emission spectrum of Eq. \ref{eq:sumharm},
$g(n,e) L_{\rm circ}$
as a function of $f=n\nu$ consists of points
along a skewed
bell-shaped curve with maximum near the relative
angular velocity at pericentre, where the greatest accelerations
are experienced ($2\pi\nu\sim \Omega_p$, where $\Omega_p$ is the angular
velocity of the relative orbit at pericentre, $v_p/r_p$).
In terms of harmonic number, a good approximation for all $e$
(becoming very good for $e>0.5$) is that $L_f$ peaks at
$n=1.63(1-e)^{-3/2}$, and $f L_f$ peaks at
$n=2.16(1-e)^{-3/2}$.

\section{Evolution to the DD stage}
\label{sec:evol}

We shall see that the GW background is dominated by the emission from
close double degenerate (DD) binaries at frequencies $10^{-4} \la f \la 10^{-1}$ Hz. In
this work, the term DD will refer to WD--WD pairs and loosely to
WD--naked helium star pairs, i.e. we exclude neutron stars from our definition. In
this section we describe the two main evolutionary pathways from the
zero-age main sequence (ZAMS) to the close DD stage. The route
followed depends mainly on the initial orbital separation of the
ZAMS stars. Similar descriptions can be found
in e.g. \cite{web98}.

We begin with an intermediate-mass ZAMS binary system with primary mass $M_1$, secondary
mass $M_2$ ($< M_1$), semi-major axis $a$ and eccentricity $e$. The
orbit may evolve somewhat due to tidal interactions between the stars,
particularly if they have convective envelopes. When the primary
evolves off the main sequence and swells in size, it may fill its
Roche lobe and start to transfer matter on to the secondary. The
stability of this mass transfer determines which of the two main
pathways to the DD stage is commenced.

\subsection{CEE+CEE}

If the primary fills its Roche lobe when it has a deep convective
envelope (i.e. on the red giant branch (RGB) or asymptotic giant
branch (AGB)), then for mass ratios $M_1/M_2 \ga 0.6$, the ensuing
mass transfer is dynamically unstable (for conservative transfer). The envelope of the primary
spills on to the secondary on a dynamical timescale, leading to the
formation of a common envelope, inside which orbit the secondary and the core of
the primary. The envelope is frictionally heated at the
expense of the stars' orbital energy, until eventually either they
coalesce, or the envelope is heated sufficiently that it is ejected from
the system, leaving the primary's core (a hot subdwarf which will
rapidly cool to become a WD, or if the primary was on
the RGB and had mass $M_1 \ga 2 \rm{~M}_\odot$, then a helium star
which will evolve to the WD stage). The basic idea of the common
envelope phase is well accepted and observationally motivated, though not well
simulated (see e.g. \citealt{liv88}; \citealt{ibe93};
\citealt{taa00}). Several formalisms have been proposed to model it in population synthesis studies. The
evolution code used here (see section \ref{sec:bse})  follows closely
the prescription of \citet{tt97} (originally from \citealt{web84}), in which
\begin{equation}
E_{\rm{bind},i}=\alpha (E_{\rm{orb},f}-E_{\rm{orb},i}),
\end{equation}
where $E_{\rm{bind,i}}$ is the initial binding energy of the envelope of the
overflowing giant star (or the sum of both envelopes' binding energies
if both stars are giants), parametrized by
$E_{\rm{bind,i}}=-G/\lambda (M_1 M_{\rm{env},1}/R_1)$, where $\lambda$
is of order unity, and is calculated in the \textsc{bse} code (see
section \ref{sec:candidates}). $E_{\rm{orb},i}$ and $E_{\rm{orb},f}$ are respectively the
initial and final orbital binding energies of the core-plus-secondary system, and
$\alpha$ is the so-called common envelope
efficiency parameter, also of order unity, usually taken to be a
parameter to be fitted to
observations. Variations to this prescription will be considered in sections \ref{sec:obs} and \ref{sec:candidates}.

Continuing with the system's evolution, the secondary star later
evolves off the main sequence, and a second common envelope phase is likely to
occur, leading to further orbital shrinkage. If once again the stars
do not coalesce then we will be left with a close(r) pair of remnants,
one or both of which may be helium stars, which in time will evolve to the
WD stage. (It is not uncommon for either helium star to overflow its
Roche lobe upon leaving the helium main sequence; this can lead to
either stable mass transfer or to a futher common envelope phase.) In
this picture, the second-formed WD will be the less massive of the
pair, since the giant star from which it descended had a smaller core
mass when its core growth was halted as it lost its envelope.
\subsection{Stable RLOF+CEE}

If Roche lobe overflow occurs when the primary is in the Hertzsprung gap, that is after the primary has exhausted its core
hydrogen and before it has developed a deep convective envelope and
ascended the giant branch, then
Roche lobe overflow may be dynamically stable for moderate mass
ratios, and a phase of stable but rapid mass transfer can occur. In this way,
the primary transfers its envelope to the secondary, leaving a compact
remnant, and a
common envelope phase is avoided, since by the time the primary
evolves to the giant branch, the mass ratio has been sufficiently
inverted that mass transfer remains dynamically stable. The
orbital separation will typically have increased during this phase
(for conservative mass transfer at least), since much of the
transfer was from the less-massive to the more-massive star. When the secondary evolves off the main
sequence, it will most likely fill its Roche lobe on the RGB, so that a common envelope phase ensues, and a
close DD is born, provided that the resulting orbital shrinkage does not lead to
coalescence. The second-formed WD will this time be the more massive,
since its progenitor was the more evolved at the time of its overflow.

The initial conditions for this route occupy a smaller range in
initial orbital semimajor axis than the CEE+CEE route, but as it results in
the injection of DD systems only at very short periods, we expect both
pathways to be significant contributors to the close DD population,
i.e. those systems contributing to the GW background in the LISA waveband.
We note also that both routes ought to
lead to the production of DDs with circular orbits, even if the ZAMS
eccentricity was non-zero, since tidal circularisation is rapid when a
system contains a near-Roche lobe-filling convective star.

\section{Analytic arguments about spectral shape}
\label{sec:anal}

Given only the above, we can make
some predictions as to the shape of the GW spectrum seen
today. A somewhat analogous treatment is given in \citet{hil90}. We
consider the evolution under GW emission of a population of DDs after creation as in
Section \ref{sec:evol}, with circular orbits. We deal here with
detached systems; the spectral shape due to interacting pairs is
discussed in section \ref{sec:inter}.

Here and throughout, we use $\nu$ for orbital frequencies and $f$ for
gravitational wave frequencies. For circular orbits, $f=2\nu$.

The number density $N(\nu,t)$ of binary WDs per unit orbital frequency
interval at time $t$ must obey the continuity equation
\begin{equation}
\frac{\partial N}{\partial t}+\frac{\partial}{\partial \nu} (\dot \nu N)=\dot N_{\rm{b}}(\nu,t),
\label{eq:cont}
\end{equation}
where $\dot N_{\rm{b}}(\nu,t)$ is the birth rate (after nuclear
evolution and mass transfer) of WD--WD systems per
unit frequency. Now for a given
source, we know that $\dot E_{\rm{orb}} = - L_{\rm{gw}}$, and using Eq.
\ref{eq:lcirc} along with $E_{\rm{orb}}=-G M_1 M_2/(2a)$ and Kepler's
law, we obtain
\begin{eqnarray}
\label{eq:fdot}
\dot \nu &=&\frac{96}{5}(2 \pi)^{8/3} \left(\frac{G {\cal
    M}}{c^3}\right)^{5/3} \nu^{11/3} \nonumber\\
&\equiv& K \nu^{11/3} \nonumber\\
&\simeq& (3.7 \times 10^{-6} \rm{ s}^{-2}) ({\cal
    M}/\rm{M_{\odot}})^{5/3} \nu^{11/3},
\end{eqnarray}
where we have used the definition of the chirp mass ${\cal M}$,
\begin{equation}
{\cal M} \equiv \frac{M_1^{3/5} M_2^{3/5}}{(M_1+M_2)^{1/5}}.
\end{equation}

We solve Eq. \ref{eq:fdot} to give the evolution $\nu(t)$
for $\dot N_{\rm{b}}=\delta(t-t',\nu-\nu')$, i.e. for a single source
injected at frequency $\nu'$ at time $t'$,

\begin{equation}
\nu(t)^{-8/3}-\nu'^{-8/3}=8K(t'-t)/3.
\end{equation}
The corresponding source number
density (Green's function for Eq. \ref{eq:cont}) $N_G(\nu,t;\nu',t')$ as a function of time is given by
\begin{eqnarray}
N_G \, d\nu & \propto & dt(\nu) \nonumber\\
& \propto & \frac{d\nu}{{\cal M}^{5/3} \nu^{11/3}} \;
 \delta\left(t-\left[t'+\frac{3}{8K}(\nu'^{-8/3}-\nu^{-8/3})\right]\right)\nonumber\\
\end{eqnarray}
since, as the system traces out a path in
$\nu$, it spends a time at each point inversely
proportional to its velocity $\dot \nu$ through frequency space.

We then consider a real injection spectrum $\dot
N_{\rm{b}}(\nu',t')$, for $\nu_{\rm{min}} < \nu' < \nu_{\rm{max}}$. The resulting number density $N(\nu,t)$ is
given by
\begin{equation}
N(\nu,t)=\int_{\nu_{\rm{min}}}^{\nu_{\rm{max}}} \int_{0}^t \dot
N_{\rm{b}}(\nu',t') \, N_G(\nu,t;\nu',t') \, dt' \, d\nu'.
\label{eq:spec}
\end{equation}
Since $L_{\rm{gw}} \propto \nu^{10/3} {\cal M}^{10/3}$, we can then
construct the GW emission
spectrum by taking $F_{\rm{gw}}(f,t) \propto f^{10/3} {\cal M}^{10/3}
N(f,t)$.

The choice of DD injection spectrum is therefore instrumental in
determining the shape of the GW emission spectrum. We can estimate its
shape as follows: we will later choose to distribute ZAMS orbital
semimajor axis uniformly in $\log a$, i.e. also uniformly in $\log
\nu$, for given initial $M_1$ and $M_2$. We suppose that,
for at least the CEE+CEE route (see section \ref{sec:evol}), the common envelope phases lead to
some mean orbital shrinkage factor, so that WD--WD pairs at their birth are also
distributed roughly uniformly in $\log \nu$. We then have $\dot
N_b(\nu') \propto 1/\nu'$, from some
$\nu_{\rm{min}} \ll \nu$
of interest, up to $\nu_{\rm{max}}$ (see also fig. 1 of \citealt{web98}). This is the maximum orbital
frequency at which a system can exit a common envelope phase and
survive to become a WD--WD pair. Upon CE exit, the newly exposed
stellar core will be a hot subdwarf, larger than the WD it will
cool to become, or it could be a naked helium star, which will
eventually evolve to the WD stage. The maximum injection frequency at WD--WD
birth is set by the minimum orbital separation that will keep this
object (and the first-formed WD) from overflowing its Roche lobe on the way to the WD
stage, whether this is at the exit of common envelope or (applicable
to the helium star case) as its radius changes due to nuclear evolution.

\begin{figure}
\includegraphics[width=80mm]{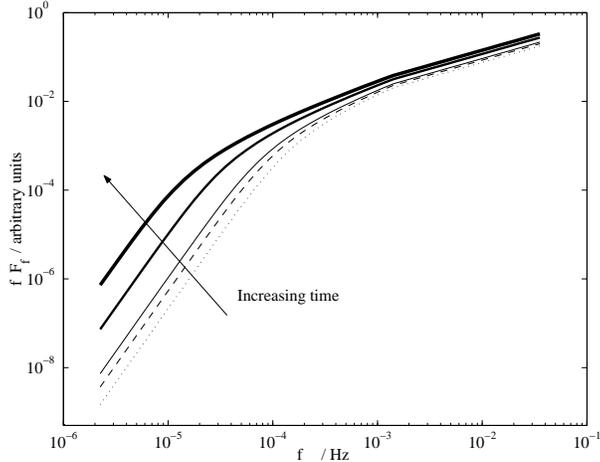}
\caption{Gravitational wave spectrum arising from constant WD--WD
  formation rate, at times 2, 5, 10, 100 and 1000 Gyr, increasing in the direction of
  the arrow shown.}
\label{fig:sfh_const}
\end{figure}

For illustration, we compute the emergent spectrum for a fiducial
population of $0.5 \rm{~M}_\odot$ WD--WD pairs. The radius of a 0.5
M$_\odot$ naked helium star does not exceed $\sim 0.13$ R$_\odot$ on its way to
the WD stage, which sets $\nu_{\rm{max}} \sim 0.7$ mHz.
If we then assign a constant pair formation rate, so that $\dot
N_b(\nu',t') = \dot N_b(\nu')$, and perform the integral in Eq. \ref{eq:spec}, we obtain the spectral shape shown in
Fig. \ref{fig:sfh_const}. Note that the spectrum is
truncated at a frequency above which the inspiralling
WDs would undergo Roche lobe overflow and merge, $f_{\rm{merge}}=2 \nu_{\rm{merge}} \simeq 40$ mHz.

\begin{figure}
\includegraphics[width=80mm]{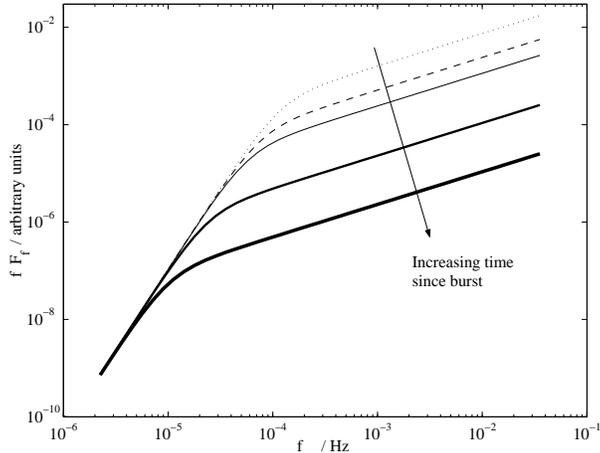}
\caption{Gravitational wave spectrum arising from a burst of WD--WD
  formation between 0 and 1 Gyr. Curves plotted are spectra at times
  2, 5, 10, 100 and 1000 Gyr, increasing in the direction of the arrow shown.}
\label{fig:sfh_burst}
\end{figure}

If instead we only inject sources for $0 < t < \tau$, and look at the
spectrum obtained for $t>\tau=1$ Gyr, (Fig. \ref{fig:sfh_burst}), we
see that the basic spectral shape is little affected.

Because of the strong dependence of $\dot \nu$ on $\nu$, a given
system of specified age will either have merged or will have remained at essentially constant
separation. Thus there are two clear physical regimes displayed in the spectra,
separated by the injection frequency from which a source could have reached
contact due to GW losses in the time $t$ since its birth, $\nu_{\rm{crit}}
\simeq 0.03-0.04$ mHz for $t \sim 5-10$ Gyr. (In all relevant
situations for us, $\nu_{\rm{crit}} < \nu_{\rm{max}}$.)  

At $f < 2\nu_{\rm{crit}}$ lies
a `static regime', in which losses due to GW are negligible in
the time available, giving $N(\nu) \propto \nu^{-1}$ and hence
$f F_{\rm{gw}} \propto f^{10/3} {\cal M}^{10/3}$. For $f \ga 2 \nu_{\rm{crit}}$, we are in the
`spiral-in' regime. In the case of a burst of DD formation (Fig. \ref{fig:sfh_burst}), sources
simply sweep through this region on the way to merger, so that we have
$N(\nu) \propto \nu^{-11/3} {\cal M}^{-5/3}$, giving $f
F_{\rm{gw}} \propto f^{2/3} {\cal M}^{5/3}$. If we have a constant DD formation rate
(Fig. \ref{fig:sfh_const}), then for $2 \nu_{\rm{crit}} < f < 2 \nu_{\rm{max}}$, merging systems are
continually being injected, so that $N(\nu)$ is less
steeply decreasing than $\nu^{-11/3}$ in this region. For $f > 2
\nu_{\rm{max}}$ the spectral slope is again $2/3$. Reality will be some
combination of these histories.

We therefore expect the cosmological spectrum we calculate later
(section \ref{sec:cosmocalc}) to be composed of a superposition of curves of these
shapes, modified for chirp mass variations, redshift effects and time
delay between progenitor star formation and DD formation. The detailed
calculations described in following sections follow in detail the evolution of all
sources from ZAMS to merger, and do not rely upon approximate
treatments of the kind given above. Simple estimates of the background
amplitude are discussed in section \ref{sec:results}.

\section{Model construction}
\label{sec:test}

\subsection{The {\sevensize\bf BSE} code and population synthesis}
\label{sec:bse}

The rapid evolution code \textsc{bse} \citep*{hur02} is used
throughout this work whenever a binary system is evolved. This code
is a fit to detailed models of stellar evolution, and produces an
evolutionary time-sequence $x(t_j)$ of the properties $x$ of any input ZAMS binary system. The
code's time-resolution adapts to the shortest current timescale for change of
the system components and orbit, due to e.g. nuclear evolution, angular
momentum loss or mass transfer, which are all treated iteratively and
have finite duration. In this way, even the most fleeting of
evolutionary phases is captured in detail, without requiring excessive
time resolution during long phases in which little changes. This is
especially useful in the study of gravitational waves, since the
majority of the GW emission from a given system occurs over an
inspiral timescale much shorter than the nuclear timescales of the
binary's parent ZAMS stars. Some of the most relevant features of the
\textsc{bse} code will be described in the following section; see
\citet{hur02} for full details.

The output $x(t_j)$ from the code can be used to construct a stellar
population at time $T$ as follows. This method is similar to that used
by \citet{hur02} to characterise the Galactic binary population.

We describe the ZAMS binary parameter
space in terms of the primary (larger) mass $M_1$, secondary mass
$M_2$ (or mass ratio $q=M_2/M_1 \le 1$), orbital
semi-major axis $a$ and orbital eccentricity $e$. We divide this space
into grid boxes, and from each box $k$, we randomly choose a ZAMS system
to represent the evolution of all sources in that box.

The number ${\cal P}_k$ of sources born into box $k$ per unit binary
system realised is determined by probability distributions $ A(a)$, $\Xi(e)$ and
$\Phi(M_1, M_2)=\Phi_1(M_1) f(q)$ in the ZAMS system properties
described above (see section \ref{sec:candidates}). ${\cal P}_k$ is obtained by
integrating the product of these distribution functions over the extent of box $k$.

We wish to construct the population of sources present at time $T$. For each output timestep $t_j$, the system with properties $x(t_j)$ can be
viewed as a system born between times $(T-t_{j+1})$ and $(T-t_j)$. If at this point the star
formation rate was ${\cal R} = {\cal R}(T-t_j)$ (expressed as a number of
binary systems born per unit time), then the number of
systems with properties $x(t_j)$ we expect to see at time $T$ is given
by
\begin{equation}
{\cal N}_{j,k}(T) = (t_{j+1}-t_{j}) {\cal R}(T-t_j) {\cal P}_k,
\label{eq:Nkj}
\end{equation}
so long as $T>t_{j}$, so that stars were not born before time began. We perform this calculation for all boxes $k$
and all timesteps $j$,
so that the total population at time $T$ is given by the combination of
all ${\cal N}_{j,k}(T)$.

This method of population synthesis ensures that sources from even
unlikely regions of ZAMS parameter space are represented,
weighted by their low formation probability. Coupled with the adaptive
time-resolution of the \textsc{bse} output, and a sufficiently fine
grid spacing, this technique allows the synthesis of a statistically
reasonable population in a modest amount of computing
time. Alternatively, statistical accuracy can be ensured with a Monte
Carlo approach by simply generating a large enough number of stars
under the initial distribution functions (see \citealt{bel02}).

Our grid extends from 0.08 to 20 M$_\odot$ in the mass of the primary
$M_1$, and from 0.08 M$_\odot$ to $M_1$ in the secondary's mass. The
initial separation is gridded from
$2(R_1+R_2)$ to $10^5(R_1+R_2)$, where $R_1$ and $R_2$ are the ZAMS radii
of the primary and secondary respectively. We
find that our background fluxes are statistically accurate to around
one per cent if we choose grid spacings of 0.05 in $\ln M$ for each
mass and 0.1 in $\ln a$ for the separation. This corresponds to
evolving $\sim 7 \times 10^5$ binaries. For the Galactic tests
described in section \ref{sec:candidates} we find that it is
sufficient to use a grid spacing twice as large in each dimension.

The \textsc{bse} code has previously been tested against various Galactic
populations of binary stars \citep{hur00}. A set of input parameters and
distributions is recommended for use with the code, to best reproduce the
observed Galactic binary population as a whole. However, in this work we are keen to
quantify the effects of astrophysical uncertainties upon population
synthesis calculations of the GW background, and so in the following
subsections we construct a set of models which differ in their
choice of input parameters but produce specifically a Galactic DD population not
in conflict with observations. The current observational uncertainties
about DDs admit a range of models. This set of models is
then considered representative of the population synthesis
uncertainties affecting the GW background.

\subsection{The state of observations}
\label{sec:obs}

The observations of DD stars are currently undergoing a
revolution. Full results of this revolution have not yet been
published, so the detailed comparison of synthesised populations with
observations is still difficult.

\citet{mar00} reported on the 15 then known DDs with measured
periods, six of which had measured component mass ratios
\citep*{max02}. Searches for DDs have mainly focussed on low-mass WDs,
$M_{\rm{WD}} \la 0.5 \rm{~M}_\odot$
(e.g. \citealt*{mar95}), since these must have formed through
giant stars losing their envelopes in binary systems, before the
helium burning that would inevitably occur in a single star. \citet{max99} determined that the fraction of DDs among
these DA WDs is between 1.7 and 19 per cent, with 95 per cent confidence.
Statistical comparisons with population synthesis models are
thus difficult, given the sample size and level of bias, but there are some
notable disagreements between observations and theory that are not
easily explained in terms of selection effects. The first of
these is the lack of observed very low mass He WDs ($M \sim 0.25
\rm{~M_\odot}$). Theory predicts an abundance of such
sources. \citet{nel01a} suggest that this can be explained by a more
rapid cooling law for low-mass WDs than is commonly used. The second discrepancy is in
the distribution of known DD mass ratios, which is seen to peak near
unity \citep{max02}. Even considering selection effects \citep{nel01a},
this is difficult to explain in terms of either standard DD formation
route, since as described in Section \ref{sec:evol}, the WD masses are
expected to differ significantly.

This prompted \citet{nel00} to suggest an alternative scenario in which a common
envelope phase between a giant and a main sequence star of similar
mass does not result in a substantial spiral-in of the
orbit, meaning that the second common envelope phase does not occur until the
secondary's radius is larger (relative to that of the primary when it
filled its own Roche lobe) than in the standard CEE+CEE picture, so that the
second WD formed is more massive, closer to the mass of the
first-formed WD. They motivate this choice by parametrizing in terms of an angular
momentum, rather than an energy balance (cf. section \ref{sec:evol}).

The observational sample of DDs is currently being substantially
increased by the SPY project \citep{nap02}, a
spectroscopic study of $\sim 1500$ apparently single WDs (not
restricted to low mass) to search for radial velocity variations
indicative of binarity. \citet{nap02} report that of the 558 WDs
surveyed so far, 90 (16 per cent) show evidence for a close WD
companion. Of these, mass
ratio determinations are reported for three DDs \citep{kar02}, these
three continuing the observed trend of mass ratios near unity.

The results of the SPY project, once analysed fully,
will help to constrain DD population synthesis calculations in a
greatly improved way. However, given the preliminary and partial
nature of the results so far, we can make only rather broad statements
about their compatibility with any given synthesised Galactic
population. This process is described in the next section.

\subsection{Candidate models}
\label{sec:candidates}

Our fiducial population synthesis model (Model A) is similar to the
preferred model suggested
by \citet{hur00} (also his Model A): we use the initial mass function (IMF) of
\citet*{kro93} (KTG) for $\Phi(M_1)$, we distribute $M_2$ uniformly in the mass
ratio $q=M_2/M_1$, $f(q)=1$, and we start with a flat
distribution in $\log a$, choosing our limits as
$2(R_1+R_2)<a<10^5(R_1+R_2)$, where $R_1$ and $R_2$ are the ZAMS radii
of the primary and secondary respectively. We have tidal effects
switched `on', we use $\alpha=3.0$ for the common envelope efficiency parameter,
and we assign all stars solar metallicity,
$Z=0.02$. For the Galaxy, we adopt the constant star formation rate
${\cal R}$ over the past 10 Gyr which gives a stellar disk mass of $6
\times 10^{10} \rm{~M_\odot}$ today.

We differ from Hurley's Model A in three main ways: first, we assign an
initial binary fraction of 50 per cent (cf. Hurley's 100 per cent) since this is observed
locally to be the case \citep{duq91} and we evolve a set of
single stars alongside the binaries, distributed according to the same
IMF as the binary primaries. Second, we assign a ZAMS
orbital eccentricity $e$ to all systems, according to a thermal
distribution $\Xi(e)=2e, 0<e<1.0$. \citet{hur00} finds that an $e=0$
model gives a somewhat better fit to observations (though he finds
that the numbers of \emph{close} ($P < 10$ d) DD systems produced are not affected); we will also test
a model of this type as part of our parameter variation (see
below). Lastly, Hurley's Model A assumed the envelope binding energy
parameter $\lambda=0.5$ for all stars, whereas here we allow this
parameter to be calculated in the code (values of $\lambda$ are from fits
to detailed models of stellar evolution by O. Pols and are an addition
to the code described in \citealt{hur02}; J. Hurley, private
communication, 2003), and in addition
we include 50 per cent of the envelope's ionisation energy in its
binding energy.

We test our synthesised Galactic populations against observations in a
necessarily simple way. The aim is to reject models in clear conflict
with the observed population of double degenerate stars, and to admit
all others as representative of the uncertainties in DD population
synthesis. Since the
overall normalisation for the cosmological integral will be entirely
separate from that used for the Galaxy, we choose primarily to compare relative
populations as opposed to absolute numbers of Galactic sources. An ideal criterion is the fraction among field WDs of close DD
binaries, which currently available SPY results place at 16 per cent. Since
the sample size is substantially larger than that of \citet{max99}, we
adopt the SPY data, despite their incompleteness. We assume a
negligible false-positive rate for SPY, and approximate the survey as
magnitude-limited ($V<15$) for the purposes of comparison. The
somewhat approximate Galactic model and star formation history used
here are sufficient, given the generosity of our selection criteria
and the fact that we compare fractional quantities
wherever possible.

We distribute all stars according to a simple double exponential
Galactic disk model
(scale height 200 pc, scale radius 2.5 kpc), then calculate the fraction of
WDs with $V<15$ expected to be members of DD binaries with
$P<100$ d. We then require that this calculated fraction be at least
10 per cent, if a given
model is to be accepted. We assign a lower limit only, since our
calculated binary fractions are likely to be overestimates, for several
reasons. First, 100 d is a generous upper limit to the orbital periods
detectable with SPY; second, we do not address the issue of the substantial
lack of observed low-mass (hence binary-member)
WDs found in other population synthesis studies; and finally, the cooling
curves used are the simple Mestel curves from \citet{hur02}; if we
instead use
the `modified Mestel cooling' from \citet{hur03}, which better fits
the theoretical curves of \citet{han99}, then our calculated binary
fraction decreases by a few percent. For our fiducial Model A, with
\citet{hur02} cooling, we find that 18 per cent of field WDs will show up as DDs in such a survey, in reasonable
agreement with the SPY results.

We also find a local total space density of WDs of $9 \times 10^{-3}
\rm{~pc}^{-3}$, and compare this with observational values, which range
from $\sim 4-20 \times 10^{-3} \rm{~pc}^{-3}$ \citep[][and
  references therein]{nel01a}. We do not attempt to compare to
distributions in mass, mass ratio or period in detail: the observed distributions are
subject to complex selection effects, and turn out often to be most
constraining for WD cooling models \citep[e.g.][]{nel01a}, whose development is beyond the scope
of this paper. We note however that
in a volume-limited sense, the mean mass ratio (where $q<1$ by
definition) for detached WD--WD pairs is $\left< q \right> = 0.62$, not
in good agreement with observations, but in common with other studies.

We then go on to consider adjustments to our model, varying the
initial distributions and mass transfer prescriptions. In all respects
other than those mentioned below, these models are identical to Model A.

In Models B, C and D, we use common envelope efficiency parameters $\alpha$ of 1.0, 2.0 and
4.0 respectively, while Model E uses the angular momentum formalism
proposed by \citet{nel00} for the first phase of spiral in, with their
recommended value of $\gamma$, and with $\alpha=4.0$.

In models N, O, P and W, we also perturb the common envelope phase. In Model N,
we include all of the envelope's ionisation energy (a positive
quantity corresponding to the energy released when the ionised
part of the envelope recombines) in its binding
energy, meaning that envelopes will be less strongly bound and hence
their removal will require less orbital shrinkage. This effect becomes
important for stars on the AGB. Model O, on the other hand, does not
include any of the ionisation energy.

Determinations of $\lambda$ from stellar modelling are found to depend on the
definition of the core-envelope boundary \citep{tau01} in giant
stars. Because of this uncertainty, we also evolve models W and P in which we fix
$\lambda=0.5$, with $\alpha=3$ and $\alpha=4$, respectively.

In Model F, we choose the primary mass from the IMF of \citet{sca86}, as in
\citet{sch01}. Then in Model G we select both $M_1$ and $M_2$ independently
from the KTG IMF, as suggested by \citet{kro93}. We also evolve a
Model K, in which  initial orbital eccentricities are set to zero.

Models L and M alter the production of DDs via the RLOF+CEE route
described in section \ref{sec:evol}. It has been suggested
\citep*{han00} that Roche lobe overflow may be stable until later in
the Hertzsprung gap (HG) than happens using the \textsc{bse} code, so a Model with
enhanced HG overflow was added (Model L). Model M has semiconservative
overflow during this stage, to emphasise the uncertainties associated
with HG mass transfer.

The Galactic DD population was simulated using each model in turn;
the results of this exercise are summarised in Table
\ref{tab:gal_models}. Imposing the criterion given above, we
eliminate Models B, G and W based on their under-production of
DDs. If we increase the binary fraction to 100 per cent, this tends
to under-produce single WDs, leading to an especially high DD fraction and
a low overall WD space density. Note that the table also contains a
Model H, which is in agreement with observations and is described in
the next section.

Thus the models A, C, D, E, F, H, K, L, M, N, O and P progress to the next round, as
representative of reasonable astrophysical uncertainties in our population
synthesis calculations. Three further models are added later (section \ref{sec:csfh});
these vary in their cosmic star formation and metallicity histories,
and so cannot be tested against the Galactic DD population.

\begin{table}
\caption{Properties of Galactic DD models; details of models given in
  section \ref{sec:candidates}.  \% DD is the percentage of field WDs in a magnitude-limited survey
  that will have a WD companion in an orbit with $P<100$
  d. $\rho_{\rm{WD},\odot}$ is the local space density of WDs (single and
  double). $\left< q \right>_{\rm{vol}}$ is the volume-limited average detached DD mass
  ratio $q$, where $q \le 1$ by definition.}
\label{tab:gal_models}
\begin{tabular}{ccccc}
\hline
Model &  \% DD & $\rho_{\rm{WD},\odot}$ & $\left< q \right>_{\rm{vol}}$
& Acceptable? \\
 & & $(10^{-3} \rm{pc}^{-3})$ & & \\
\hline
A & 18 & 9 & 0.62 & Yes\\
B & 7 & 8 & 0.68 & No\\
C & 13 & 9 & 0.63 & Yes\\
D & 20 & 9 & 0.63 & Yes\\
E & 24 & 9 & 0.75 & Yes\\
F & 22 & 6 & 0.64 & Yes\\
G & 6 & 6 & 0.58 & No\\
H & 18 & 9 & 0.62 & Yes\\
K & 17 & 9 & 0.63 & Yes\\
L & 18 & 9 & 0.63 & Yes\\
M & 17 & 9 & 0.62 & Yes\\
N & 17 & 9 & 0.63 & Yes\\
O & 20 & 9 & 0.59 & Yes\\
W & 9 & 8 & 0.62 & No\\
P & 12 & 8 & 0.62 & Yes\\
\hline
\end{tabular}
\end{table}

\subsection{Interacting DDs and modifications made to BSE code}
\label{sec:alter}

Some modifications were made to the \textsc{bse} code regarding the treatment of accreting DD systems. In this we
mainly follow the recommendations made in the detailed population
synthesis work of \citet{nel01b}.

AM CVn stars are
mass-transferring compact binaries in which the transfer is driven by
gravitational radiation, and in which the
accretor is a white dwarf and the donor is a Roche-lobe filling star,
which could be another (less massive) white dwarf, or a helium
star. For a review, see \citet{nel01b} and references therein. While
not expected to be the dominant source of the Galactic gravitational
wave background (\citealt{hil98}; \citealt{hb00}), some of these systems will be useful as
`verification' sources for LISA, with large, predictable
gravitational wave amplitudes.

We include in our definition of AM CVns all systems in which
a helium star or WD is transferring mass on to a WD, including those
systems in which the donor star is a CO or ONe WD.

\paragraph{The WD family}
When the donor star is a white dwarf, the orbital separation at
initial Roche lobe overflow is around 0.1 R$_{\odot}$, which is often
sufficiently small that the accretion stream impacts directly on the
accretor's surface, so an accretion disc is not expected to form. This has
implications for the orbital evolution of the mass-transferring
binary. When an accretion disc is present, tidal torques on the outer
edge of the disc return to the orbit the angular momentum carried away from
the donor by the accretion stream. In the absence of such a mechanism
for restoring the orbital angular momentum, the criterion for stable mass transfer
becomes much more stringent, and in most cases an AM CVn star will not
form, precluding the existence of the WD family. Here we take the optimistic view (as in model II
of \citealt{nel01b}) that, even if no disc is present, some
tidal mechanism has an equivalent effect and that all WD--WD systems
for which the mass ratio is $<0.628$ \citep{hur02} will commence
stable mass transfer upon Roche lobe overflow. We modify the \textsc{bse} code
accordingly. This optimism is perhaps warranted, since we \emph{do} see WD
family AM CVn systems, e.g. \citet{isr02},
which reports on the discovery of a helium-transferring
compact binary with orbital period (321 s) too short to involve a
(non-degenerate) helium star donor.

\paragraph{The helium star family}

In this case, the donor star is a helium star, produced when a star
with mass $\ga 2 \rm{~M}_\odot$ loses its envelope on the RGB.
Since these stars can live for a
rather long time compared with the main sequence lifetimes of their
progenitors, there is a significant chance that through GW losses (or
sometimes radial evolution) they will commence mass transfer before
evolution to the WD stage. Here we shall employ the same condition on
the dynamical stability of this mass transfer as \citet{nel01b}:
$q = M_{\rm{nHe}}/M_{\rm{wd}} < 1.2$ (we use `nHe' to denote (naked)
helium star, to avoid confusion with helium-core WDs). Stellar modelling
\citep*{sav86} indicates that rapid mass transfer forces the helium star out of
thermal equilibrium, increasing the thermal timescale beyond a
Hubble time. The star cannot
ever regain thermal equilibrium, and becomes semi-degenerate (as
opposed to fully degenerate) as its
mass falls. This
results in a negative exponent in the mass-radius relation, 
so that the orbital separation then increases as the helium star stably loses mass,
i.e. an AM CVn system is formed. Note that at the onset of
Roche lobe overflow, helium stars are always large enough that an
accretion disk can form.

The standard \textsc{bse} code does not incorporate the possibility of these
semi-degenerate helium stars, so this was added. Here we adopt the same
semi-degenerate mass-radius relation as in
\citet{nel01b} (in solar units):
\begin{equation}
R_{\rm{nHe}} = 0.043 \, M_{\rm{nHe}}^{-0.062},
\end{equation}
and switch between this and the regular non-degenerate relation by
selecting the larger of the two radii when the helium star is
transferring mass on to a WD companion. In our code, this changeover
occurs at $M_{\rm{nHe}} \sim 0.29 \rm{~M}_{\sun}$. We also modify the mass transfer rate prescription in the code, in
order that the transfer responds more quickly to the initial overflow,
so that the helium star does not hugely overhang its Roche lobe, and we
halt further helium burning, so that the star cannot evolve to the WD
stage during transfer, due to its long thermal timescale. We note that
this modification is fairly crude, but ought to give a good indication
of the relative importance of helium star AM CVn systems as sources of the
GW background.

A further issue in the formation of any helium-transferring system is
that of edge-lit detonations (ELDs), which are believed to occur after
a layer of helium has built up in the surface of an accreting CO
WD. The \textsc{bse} code detonates CO WDs in this way after the
accretion of 0.15 M$_\odot$ of helium. We evolve separately a model
(Model H) in which this is increased to 0.3 M$_\odot$, as in Model II of
\citet{nel01b}.

\section{Cosmological equations}
\label{sec:cosmocalc}

In this section we describe our calculation of the cosmological
background. We adopt a standard
lambda-cosmology, with $\Omega_{\rm{m}}=0.3$, $\Omega_\Lambda=0.7$ and
\mbox{$H_0=70$ km s$^{-1}$ Mpc$^{-1}$}. This means that the current
age of the universe, $T_0 = 13.5$ Gyr. We assume isotropy throughout;
for an analysis of the small anisotropy due to the localisation of binary stars
in galaxies which follow the large scale structure of the universe, see
\citet{kos00}.

\subsection{Basic Equations}

The specific flux $F_{f_{\rm{r}}} = dF(f_{\rm{r}})/df_{\rm{r}}$
  received at frequency $f_{\rm{r}}$ from an object at
  redshift $z$ with specific luminosity $L_{f_{\rm{e}}}$ is given
  by \citep[e.g.][]{pc99}

\begin{equation} F_{f_{\rm{r}}} = \frac{L_{f_{\rm{e}}}}{4 \pi d_{\rm{L}}(z)^2} \left(\frac{df_{\rm{e}}}{df_{\rm{r}}}\right), \end{equation}
where  $f_{\rm{e}}=(1+z)f_{\rm{r}}$, $d_{\rm{L}}(z) = (1+z)d_{\rm{M}}(z)$ is the
luminosity distance to
redshift $z$ and $d_{\rm{M}}$ is the proper motion distance (cf. section 5
of \citet{hg00}, which is also $1/(2\pi)$ times the proper (`comoving')
circumference of the sphere about the source which passes through the
earth today).

If the radiation comes from a large number of sources spread over
redshift and isotropically distributed on the sky, we can write $dL_{f_{\rm{e}}}(z) = \ell_{f_{\rm{e}}}(z) dV(z)$,
where $\ell_{f_{\rm{e}}}(z)$ is the comoving specific luminosity
density (say in erg s$^{-1}$ Hz$^{-1}$ Mpc$^{-3}$),
$dV(z) = 4\pi d_{\rm{M}}^2 d\chi$ is the comoving volume element and
$\chi$ is the comoving distance.

We can then write the specific flux received in gravitational waves today as
\begin{eqnarray} F_{f_{\rm{\rm{r}}}} &=& 
\int^{\infty}_{z=0} \frac{\ell_{f_{\rm{e}}}}{4\pi
  d_{\rm{L}}^2(z)}\left(\frac{df_{\rm{e}}}{df_{\rm{r}}}\right) dV(z)
\\
&=&\int^{T_0}_{T=0}
\frac{\ell_{f_{\rm{e}}}(T)}{(1+z(T))}
\left(\frac{df_{\rm{e}}}{df_{\rm{r}}}\right) c \, dT,
\label{eq:fexact}\end{eqnarray}
using $d\chi = -(1+z)c \, dT$, where $T$ is cosmic time.

This is the basic equation on which the code is based. The equation is
discretised in $f_{\rm{r}}$, $T$ and $\ell$ as described in
section \ref{sec:comp}.

\subsection{Computational Equations}
\label{sec:comp}

In the code, we bin the received gravitational waves in frequency. To calculate the flux received in a frequency bin with limits
$f_{\rm{r}1}$ and $f_{\rm{r}2}$, we integrate Eq. \ref{eq:fexact} between these limits:
\begin{eqnarray}
F_{f_{\rm{r}1}\rightarrow f_{\rm{r}2}} &=&\int_{f_{\rm{r}1}}^{f_{\rm{r}2}} \int^{T_0}_{T=0}
\frac{\ell_{f_{e}}(T)}{(1+z(T))}
\left(\frac{df_{e}}{df_{r}}\right) cdT\, df_{r} \nonumber\\
  &=& \int^{T_0}_{T=0} \int_{(1+z)f_{\rm{r}1}}^{(1+z)f_{\rm{r}2}}
\frac{\ell_{f_{e}}(T)}{(1+z(T))}
\, df_{e}\, cdT,
\end{eqnarray}
i.e. we integrate only over those emitted frequencies that will have
been redshifted to arrive in this frequency bin today. The bin size was chosen to
be $0.1$ in $\log_{10}(f_{\rm{r}})$.

Clearly, to calculate $F$, we need to know the comoving luminosity
density $\ell_{f_{e}}$ in gravitational radiation at frequency
$f_{\rm{e}}$ as a function of cosmic time.

We first obtain the source population at a given cosmic time $T_i$, by
simply generalising Eq. \ref{eq:Nkj}, so that now
\begin{equation}
N_{k,j}(T_i)=(t_{j+1}-t_j){\cal R}_c(T_i-t_j) {\cal P}_k,
\end{equation}
where ${\cal R}_c(T)$ is the cosmic star formation rate at time $T$, expressed as
a number of binary stars born per unit time per unit volume, and $N_{k,j}(T_i)$ is the number density
of binaries with parameters $k,j$ at cosmic time $T_i$, and where we
require $T_i \ge t_j$.

The gravitational wave luminosity density at time $T_i$ is then given
by
\begin{equation}
\ell_{f_{\rm{e}}}(T_i) = \sum_{k,j} N_{k,j}(T_i) L_{k,j}(f_{\rm{e}}),
\end{equation}
i.e. we simply sum over the emission at frequency $f_{\rm{e}}$ from all
sources $k,j$ present at that time, weighted by their space densities.

Since each binary source $s$ emits radiation at only specific
frequencies $f_n=n \nu_s$ (where $\nu_s$ is the orbital frequency of
binary $s$) at a given time (Eq. \ref{eq:sumharm}), this sum can be
expressed as
\begin{equation} \ell_{f_{\rm{e}}}(T_i) = \sum_{k,j} N_{k,j}(T_i)
  \sum_n L_{\rm{circ},k,j} \, g(n,e_{k,j})
  \delta(f_{\rm{e}}-n \nu_{k,j}).
\end{equation}

We then have
\begin{equation} F_{f_{\rm{r}1}\rightarrow f_{\rm{r}2}} = \sum_{i} \sum_{k,j} \sum_{n_{\rm{min}}}^{n_{\rm{max}}}
\frac{N_{k,j}(T_i) L_{\rm{circ},k,j} \, g(n,e_{k,j})}{(1+z_i)}
\;  c \Delta T, \label{eq:fcomp}\end{equation}
where we have also discretised the integral over cosmic time $T$, as a
sum over $i$ intervals $\Delta T$, and
where $n$ is an integer, with the limits $n_{\rm{min}}$ and
$n_{\rm{max}}$ defined by $f_{\rm{r}1}<\frac{n \nu_{j,k}}{1+z_i}<f_{\rm{r}2}$.
At a given redshift $z_i(T_i)$, we just sum over those
harmonics of those sources that will lead to emission at frequencies
$f_{\rm{e}}$, with $f_{\rm{r}1}(1+z_i) < f_{\rm{e}} < f_{\rm{r}2}(1+z_i)$, and hence
reception in the $f_{\rm{r}1}\rightarrow f_{\rm{r}2}$ frequency bin today.

The integration timestep $\Delta T$ must be sufficiently small
that the emitting source population does not change significantly on
timescales shorter than this, i.e. we assume a quasi-steady state population
during this interval, so that our snapshot of the population at time
$T_i$ is representative of the whole timestep $\Delta T$. A value of $\Delta T=T_0/50$ was used
throughout. We checked that timesteps smaller than this did not
yield noticeably different results. Individual sources may evolve
significantly within this timestep, but the characteristic emission of the
population will be unchanged. It should also be noted that the
evolutionary timesteps taken for the binary stars are independent of
this integration timestep (see section \ref{sec:bse}), so that
$\Delta T$ may be made much larger than the timescales of the
evolutionary processes of interest, so long as the population is
roughly steady-state over $\Delta T$.

Equation \ref{eq:fcomp} is the sum performed by the code written for
this paper, for a
large number of received frequency bins over the range
$10^{-6} < f_{\rm{r}} < 10^{0}$ Hz. For practical purposes, the
sum over harmonics is truncated when $g(n,e)$ drops below $10^{-3}$,
well beyond the peak in the emitted spectrum at
$n_p=1.63(1-e)^{-3/2}$. For  typical $e<0.95$, our numerical cutoff at
$g=10^{-3}$ corresponds roughly to including only $n<5n_p$. The higher
values $n>5n_p$ contribute less than 1 per cent of the total gravitational wave
luminosity.

\subsection{Quantities Used}
\label{sec:quantities}

Some quantities commonly used in gravitational wave astronomy are:
$F_{f_{\rm{r}}}(f_{\rm{r}})$, $4 \pi$ times the specific intensity;
$\Omega_{\rm{gw}}(f_{\rm{r}})$, the fraction of closure density per
logarithmic GW frequency interval; and the power spectral density $S_h(f_{\rm{r}})$.

The first of these, $F_{f_{\rm{r}}}(f_{\rm{r}})$, can be calculated from
\begin{equation} F_{f_{\rm{r}}}(\overline{f_{12}}) = \frac{F_{f_{\rm{r}1}\rightarrow f_{\rm{r}2}}}{(f_{\rm{r}2}-f_{\rm{r}1})}. \end{equation}

The second, $\Omega_{gw}(f_{\rm{r}})$, is the fraction of closure energy
density contained in gravitational waves received in the logarithmic
frequency interval around $f_{\rm{r}}$, i.e.

\begin{equation} \Omega_{\rm{gw}}(f_{\rm{r}}) = \frac{1}{\rho_c c^2} \frac{f_{\rm{r}}
  F_{f}(f_{\rm{r}})}{c}, \end{equation}
where $\rho_c$ is the critical mass density of the Universe;
$\rho_c = (3H_0^2/8 \pi G) \simeq (1.88 \times 10^{-29}) \, h_{100}^2$ g cm$^{-3}$, where $H_0 =
100 \, h_{100}$ km s$^{-1}$ Mpc$^{-1}$. In terms of computational quantities,
\begin{eqnarray} \Omega_{\rm{gw}}(\overline{f_{12}}) &=&\frac{1}{c^3 \rho_c}
\frac{F_{f_{\rm{r}1}\rightarrow f_{\rm{r}2}}}{\Delta (\ln f_{\rm{r}})} \nonumber\\
&\simeq&
0.0175 \; F_{f_{\rm{r}1}\rightarrow f_{\rm{r}2}} \end{eqnarray}
(where $F_{f_{\rm{r}1}\rightarrow f_{\rm{r}2}}$ is in erg s$^{-1}$ cm$^{-2}$)
since $\Delta( \ln f_{\rm{r}}) = \ln(10^{0.1}) \simeq 0.23$ and $h_{100}=0.7$.

The power spectral density $S_h(f_{\rm{r}})$ is given by
\begin{equation} S_h(f_{\rm{r}}) = \frac{4 G}{\pi c^3} \frac{1}{f_{\rm{r}}^2}
  F_{f_{\rm{r}}}(f_{\rm{r}}). \end{equation}
Usually this is plotted as $S_h^{1/2} \simeq (5.6 \times 10^{-20})
\frac{F_{f_{\rm{r}}}^{1/2}}{f_{\rm{r}}} \rm{Hz}^{-1/2}$, where $F_{f_{\rm{r}}}$ is in erg s$^{-1}$
cm$^{-2}$ Hz$^{-1}$, and $f_{\rm{r}}$ is in Hz.

\subsection{Cosmic star formation history}
\label{sec:csfh}

As pointed out by \citet{sch01}, most determinations of cosmic star formation
history are based on the UV emission from massive stars (e.g. \citealt{mad96};
\citealt{ste99}), and use an assumed single-star IMF (commonly that of \citealt{sal55}) to convert
observed UV flux into a star formation rate as a function of
redshift. This type of rate is inconvenient here for two reasons:
first, a non-trivial factor is required for conversion to a \emph{binary}
star formation rate (for an assumed binary fraction), because of the need to correct for the observed
flux from companion stars; and second, the total star formation rate is
pivoted on the high-mass end of the stellar distribution, while here we
are interested in studying the remnants of low- to intermediate-mass
stars. This results in a crucial dependence on the choice of stellar IMF.

\citet{sch01} overcome the first problem by assuming the measured
shape of the cosmic SFH as a function of time, but normalising its
amplitude to the local rate of core-collapse
supernovae. This Type Ibc/II SN rate is a more easily calculated
quantity for a given (binary or single) IMF than is the UV luminosity density. Since \citet{sch01} are also concerned with neutron stars in their study, this is
a reasonable choice. However, the second problem remains when one is
concerned with WDs; and in addition, not only does
the normalisation pivot on the high mass stars, but it also depends
crucially on the ratio of local to peak cosmic SFR. We also
note that the minimum mass of star producing a core-collapse supernova
explosion is uncertain \citep[e.g.][]{jef97}.

For our normalisation, we use instead the observed local stellar mass
density $\Omega_*$, as
derived from the local near-IR luminosity function by
\citet{col01}. This quantity is most sensitive to stellar masses near
the MS turnoff in old populations, $M \sim 0.8-1.0 \rm{~M}_\odot$, and
thus is more closely related than the SNIbc/II rate to the DD progenitor population. We convert
between their assumed single star IMF and our binary star IMFs by keeping
constant the mass in stars in this range. We then use the recycled
fraction $R=0.42$, as for the \citet{ken83} IMF used in \citet{col01}, to convert stellar density today to total mass of stars ever
formed, $\Omega_{*,\rm{tot}}$ (the time-integral of the cosmic star
formation rate). Doing
this, we obtain $\Omega_{*,\rm{tot}} = 5.0 \times 10^{8} \, \rm{M}_\odot \rm{Mpc}^{-3}$ for
the KTG IMF, while for the Scalo IMF this figure is $\Omega_{*,\rm{tot}} = 4.0 \times 10^8 \,
\rm {M}_\odot \rm{Mpc}^{-3}$. Due to this rather crude conversion, the
uncertainty in these figures will be greater than the 15 per cent quoted by
\citet{col01} for $\Omega_*$; we estimate the resulting uncertainty to
be $\sim$ 30 per cent.

\begin{figure}
\includegraphics[width=80mm]{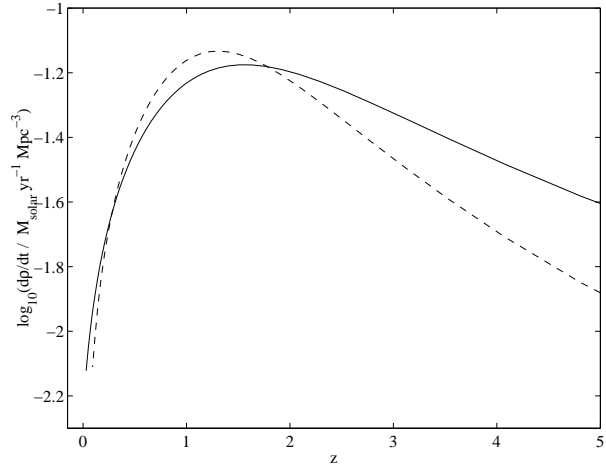}
\caption{Two possible cosmic star formation histories, plotted as a
  function of redshift $z$ and parametrized according
  to the smooth curve fits given in \citet{col01}. Dashed line: no
  extinction correction made, used in Model J. Solid line: extinction
  corrected with $E(B-V)=0.15$, used in all other Models. The time integral of each rate is fixed
  using the appropriate $\Omega_{*,\rm{tot}}$ derived from
  \citet{col01}. Curves shown are for the KTG IMF.}

\label{fig:sfh_versions}
\end{figure}

\citet{col01} note that their calculated stellar densities are most
consistent with UV-derived star formation rates if the extinction corrections used in
these methods are moderate. However, we would like to assess the effects of uncertainty in the shape
of the cosmic star formation history. We therefore select both a history
with large extinction corrections and one with none, keeping the
integral over time fixed to $\Omega_{*,\rm{tot}}$ for each. The corresponding
curves are plotted in Fig. \ref{fig:sfh_versions}. We use the
extinction-corrected rate, favoured by \citet{ste99}, in Model A and all
other models except for Model J, which uses the uncorrected rate (but is
identical to Model A in all other respects). We also introduce Models Q and R, whose metallicity histories differ from that of Model A: in Model Q, stars born during the first Gyr have metallicity $1/20$ solar, while stars born later have solar composition; in Model R, all stars have metallicity $Z=0.01$, i.e. half-solar.

\section{Basic results}
\label{sec:results}

\begin{figure}
\includegraphics[width=80mm]{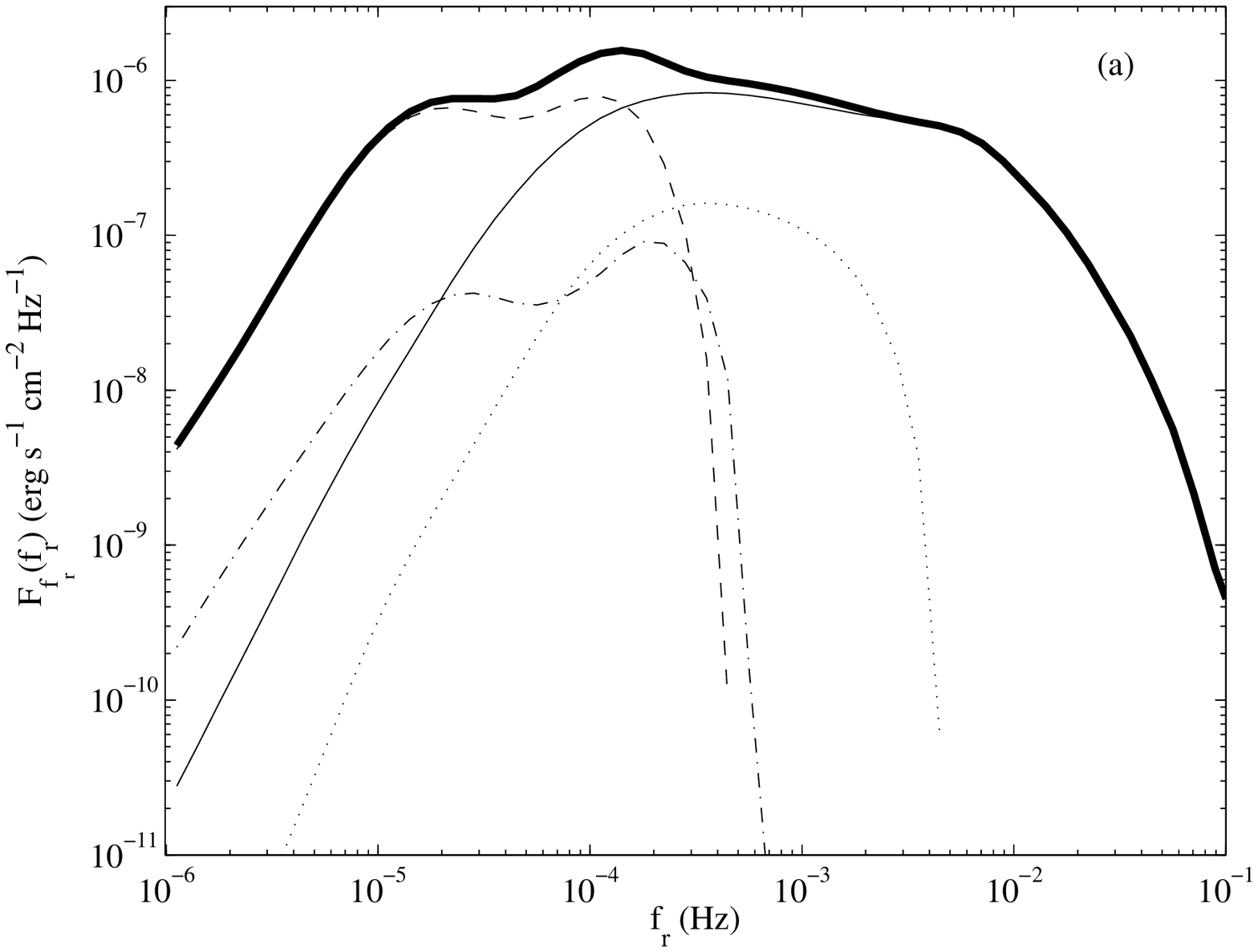}
\includegraphics[width=80mm]{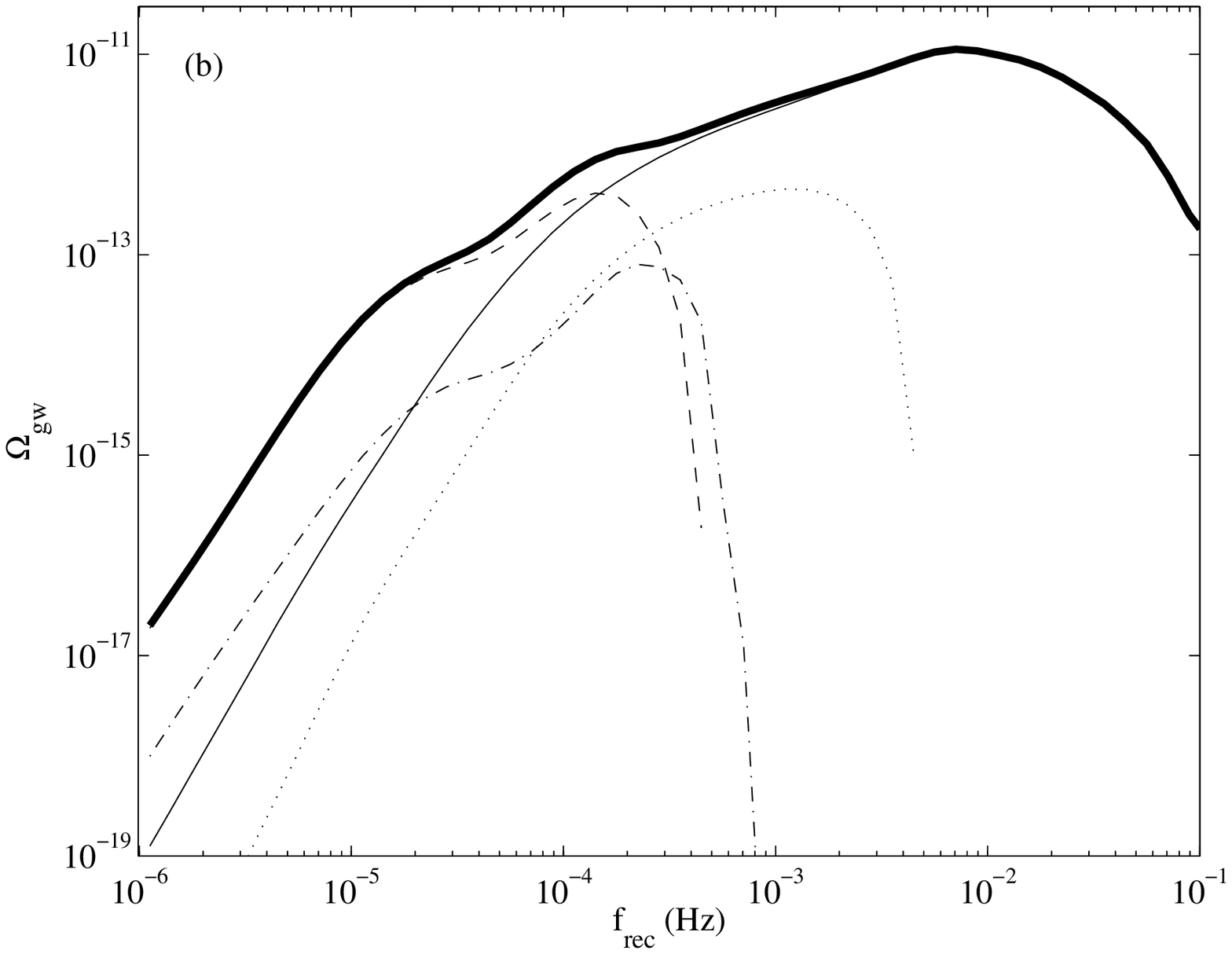}
\includegraphics[width=80mm]{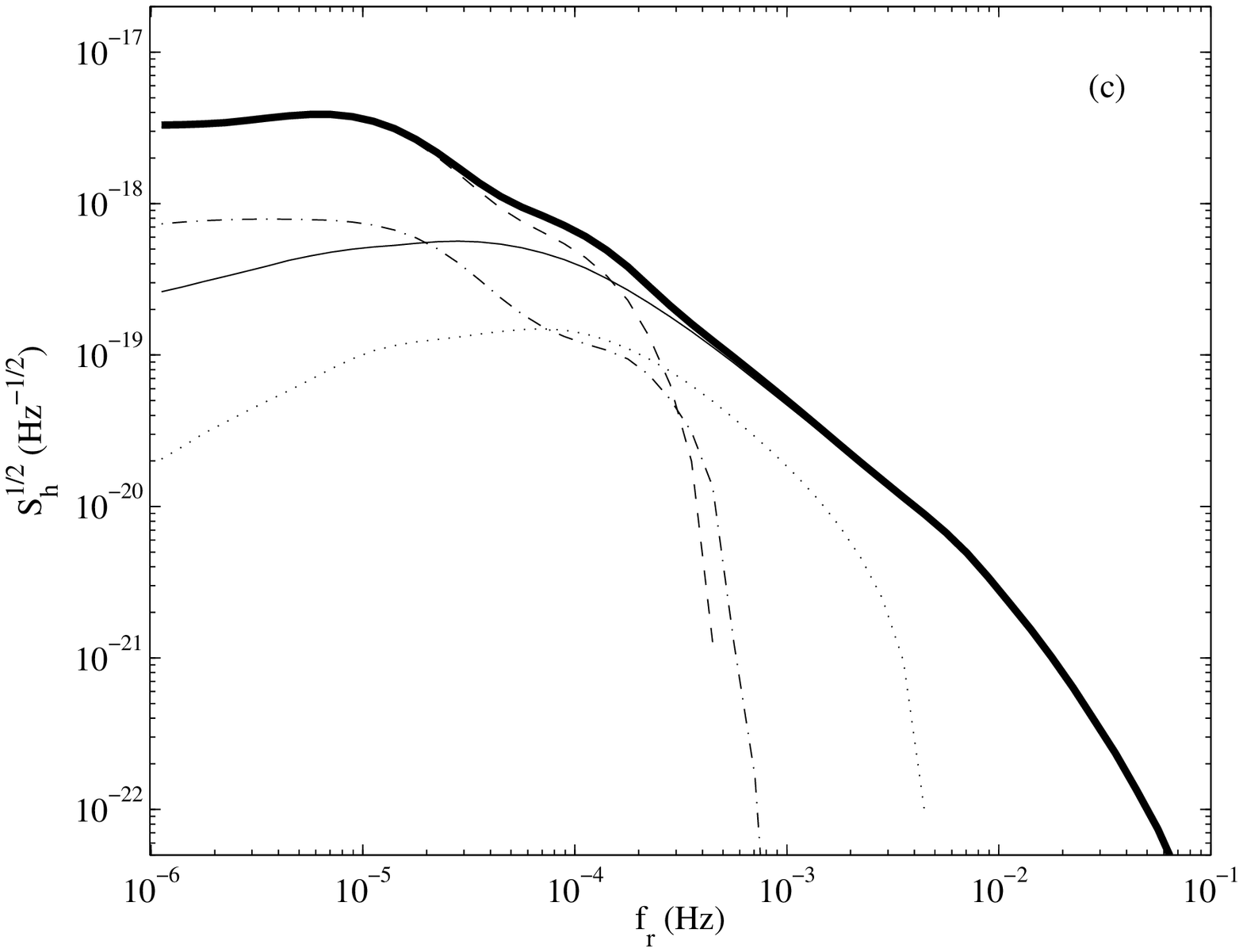}
\caption{The GW background for our fiducial Model A, in terms of the
  three quantities described in section \ref{sec:quantities}. Solid line: WD--WD pairs;
  dotted line: nHe--WD pairs; dashed line: MS--MS binaries, and
  dot-dash line: WD--MS binaries. The total signal (the sum of the four
  parts) is given by the thick solid line. Only $n=2$ harmonics of the orbital
  frequency are plotted (see section \ref{sec:ecc_test}).}
\label{fig:basic}
\end{figure}

The GW background spectrum received in the frequency range $10^{-6} < f_{\rm{r}} <
10^{-1}$ Hz, generated using our fiducial Model A, is
plotted in Fig. \ref{fig:basic}. The total amplitude is broken down
into separate contributions from four main evolutionary stages: main
sequence--main sequence (MS--MS), WD--MS, WD--WD and WD--helium star
(WD--nHe) binaries, and plotted in terms of each of $F_{f_{\rm{r}}}$, $\Omega_{\rm{gw}}$
and $S_h^{1/2}$ described in section \ref{sec:quantities}. The
unitless $\Omega_{\rm{gw}}$ will be our preferred quantity for the
remainder of the paper\footnote{Note that since \citet{col01} quote
  $\Omega_* h$ in their paper, and we use this quantity to
  normalise our star formation rate, our calculated $\Omega_{\rm{gw}}$ also scales as $h^{-1}$. We use $h=0.7$.)}.

\begin{figure}
\includegraphics[width=80mm]{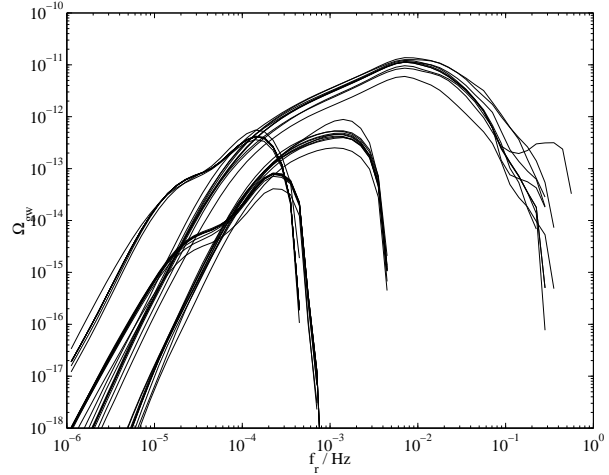}
\caption{Comparison of spectral shapes for all Models. Curves are the
  same as in Fig. \ref{fig:basic}, but all quantities are plotted as
  solid lines. Only $n=2$ harmonics of the orbital
  frequency are plotted (see section \ref{sec:ecc_test}).}
\label{fig:allmodels}
\end{figure}

The four component spectra are plotted in Fig. \ref{fig:allmodels}
for all of the models evolved, to illustrate that the spectral shapes
are largely unaffected by any of the changes made. A summary of
important quantities for each model (to be discussed later) is given
in Table \ref{tab:summary}. For reference, we also list a Model A$'$,
identical in parameters to Model A, to demonstrate the typical level
of statistical variation in the results. This is clearly at the 1 per
cent level in flux, so that variations larger than this between models
can be ascribed to parameter, and not statistical, variations.

Throughout we will focus on the properties of
the spectrum around 1 mHz, in the centre of the LISA band and of the
spiral-in regime. We will also compare with the spectral
properties at 10 mHz, at which frequency lower-mass WD--WD pairs
can no longer be present and at which point this extragalactic
WD--WD background will be the dominant LISA background source (see
Fig. \ref{fig:lisa}).

\begin{table}
\caption{Summary table of results from all models described in the
  text. $\Omega_{\rm{gw}}$(1 mHz) is in units of $10^{-12}$. R+C
  refers to the RLOF+CEE route to the DD stage, while C+C refers to
  the CEE+CEE route (see section \ref{sec:evol}; note that for
  Model E, we hold the RLOF+CEE contribution fixed from Model A). The flux-weighted mean
  chirp mass $\left< {\cal M} \right>$ contributing at $f_{\rm{r}}
  \sim$ 1 mHz is in units of M$_\odot$, the next column lists the percentage contribution to
  $\Omega_{\rm{gw}}$ at 1 mHz from interacting binaries and the last
  column gives the inspiral remnant density $N_0$ today, in units of
  $10^6$ Mpc$^{-3}$. Models B and G were rejected for reasons noted in
  Table \ref{tab:gal_models}.}
\label{tab:summary}
\begin{tabular}{ccccccc}
\hline
Model & \multicolumn{3}{c}{$\Omega_{\rm{gw}}$(1 mHz)} & $\left< {\cal
  M} \right>$ & \% & $N_0$\\
& Total & R+C & C+C & & AM CVn & \\
\hline
A & 3.57 & 1.35 & 2.22 & 0.45 & 13 & 1.17\\
A$'$ & 3.61 & 1.36 & 2.26 & 0.45 & 14 & 1.18\\
C & 3.06 & 0.60 & 2.47 & 0.44 & 16 & 0.90\\
D & 3.66 & 1.64 & 2.02 & 0.47 & 13 & 1.20\\
E & 4.21 & 1.35 & 2.86 & 0.47 & 10 & 1.57\\
F & 1.94 & 0.72 & 1.22 & 0.41 & 13 & 0.75\\
H & 4.10 & 1.53 & 2.58 & 0.43 & 25 & 1.17\\
J & 3.62 & 1.38 & 2.24 & 0.45 & 13 & 1.17\\
K & 4.29 & 2.09 & 2.20 & 0.48 & 13 & 1.29\\
L & 3.80 & 1.53 & 2.27 & 0.45 & 12 & 1.25\\
M & 2.80 & 0.66 & 2.14 & 0.44 & 15 & 0.92\\
N & 3.43 & 1.36 & 2.07 & 0.46 & 13 & 1.13\\
O & 3.89 & 1.31 & 2.57 & 0.46 & 13 & 1.27\\ 
P & 5.46 & 1.00 & 4.46 & 0.55 & 16 & 1.20\\
Q & 3.73 & 1.43 & 2.30 & 0.44 & 13 & 1.32\\
R & 3.83 & 1.48 & 2.35 & 0.44 & 12 & 1.28\\
\hline
\end{tabular}
\end{table}

It is clear that the signal in the LISA frequency band ($0.1 \la
f_{\rm{r}} \la 10$ mHz) is dominated by the WD--WD component, as
expected. Neither the MS--MS nor the MS--WD binaries can radiate at
frequencies above the bottom of this bandpass, since even
the lowest mass MS stars come into contact at frequencies below 1 mHz.
 WD--nHe pairs can contribute to a somewhat
higher frequency due to the smaller radii of helium stars, but still
come into Roche lobe contact at $f_{\rm{e}} \sim$ 1 mHz.

The WD--WD component clearly displays the spectral shape predicted in
Section \ref{sec:anal} ($\Omega_{\rm{gw}} \propto f_{\rm{r}} F_{f_{\rm{r}}}$, plotted
in Figs. \ref{fig:sfh_const} and \ref{fig:sfh_burst}), with a
clear separation between static and spiral-in regimes at around
$10^{-4}$ Hz. The slope in
the static regime suggests that sources are injected with a spectrum
closer to $\dot N_{\rm{b}}(\nu) \propto \nu^{-2/3}$ than to $\nu^{-1}$,
but agreement to this level is encouraging. The spiral-in slope is
slightly steeper than predicted, but this is due to the spectrum seen
being the sum of spectra from populations with different chirp masses,
as well as different merger and maximum injection frequencies (see
Fig. \ref{fig:wdtypes}), whose individual slopes in the spiral-in
regime are closer to the predicted 2/3. Agreement with our simple
predictions is therefore good and we feel that we understand well the
origins of the spectrum.

\subsection{Contributors}

The breakdown of contributions to the background received at 1 mHz for
our fiducial Model A is given in Table \ref{tab:wdtypes}. In this
section we identify the dominant source types, and
those types whose contribution is negligible, then attempt to characterise the
emitting population in terms of a mean chirp mass and inspiral remnant
density.

\begin{table}
\caption{Percentage contribution to $\Omega_{\rm{gw}}$ at 1 mHz from
  different DD pairs, for both contribution to total integrated
  background and contribution to background coming from the local
  universe, $z=0$. All for fiducial Model A. MS--MS and WD--MS
  binaries contribute negligibly at this frequency. (`nHe' denotes
  a naked nondegenerate helium star.) All contributions, including the
  AM CVn values, are given as fractions of the total flux at 1 mHz.}
\label{tab:wdtypes}
\begin{tabular}{lrr}
\hline
Pairing & \% over all time & \% locally\\
\hline
He--He & 12.4 & 29.5\\
He--CO & 23.0 & 25.3\\
He--ONe & 0.6 & 0.6\\
CO--CO & 42.2 & 33.2\\
CO--ONe & 8.1 & 4.4\\
ONe--ONe & 1.0 & 0.2\\
(of which AM CVn) & 3.6 & 4.7\\
\hline
nHe--WD & 12.7 & 6.9\\
(of which AM CVn) & 9.7 & 2.0\\
\hline
Total & 100 & 100\\
(of which AM CVn) & 13.3 & 6.7\\
\hline
\end{tabular}
\end{table}

\subsubsection{Eccentric harmonics}
\label{sec:ecc_test}

As described in section \ref{sec:lgw}, systems with eccentric orbits emit
gravitational waves at all harmonics $n \nu$ of the orbital frequency, not
just the $n=2$ harmonic as for circular orbits.

\begin{figure}
\includegraphics[width=80mm]{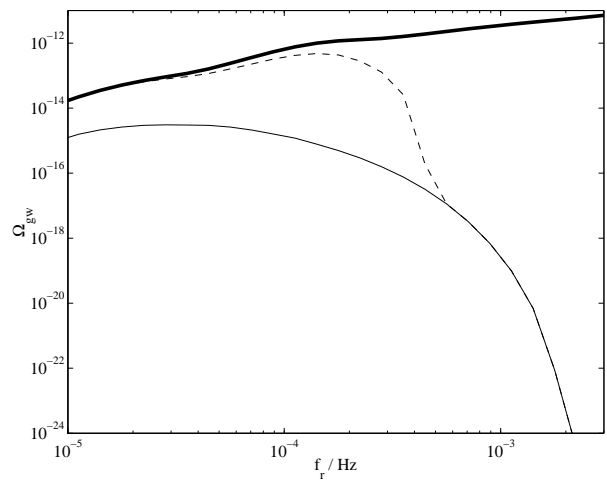}
\caption{The GW background from harmonics with $n \ge 3$ from MS--MS
  pairs (thin solid line), plotted along with the total MS--MS pair
  contribution (dashed line), and the total background from all sources
  (thick solid line), demonstrating that the harmonic
  contribution is negligible. All for Model A.}
\label{fig:harm}
\end{figure}

The only \emph{close}
binaries we expect to be eccentric are unevolved MS--MS binaries in
which tidal forces have not yet circularised the orbit. Almost every
close evolved (e.g. WD--MS, WD--WD) system will have at some point
experienced a Roche lobe-filling phase, which will likely have circularised
the system, through tidal circularisation and/or common envelope
evolution. Figure \ref{fig:harm} shows the contribution from harmonics
with $n \ge 3$ to the MS--MS GW spectrum for Model A (which has a
thermal initial eccentricity distribution). Clearly the $n \ge 3$
harmonics contribute $\la 10$ per cent of the MS--MS spectrum at frequencies
$f_{\rm{r}} \la 0.5 \rm{~mHz}$, and although they dominate the
MS--MS spectrum above this frequency, these signals are buried deep
below the other contributors at $f_{\rm{r}} > 0.5 \rm{~mHz}$ (see Fig. \ref{fig:basic}). Hereafter we safely
neglect the $n \ne 2$ contributions to $L_{\rm{gw}}$, in the
interests of computing time, though we do not neglect eccentric orbits
in computing stellar evolution sequences.

\subsubsection{Interacting binaries}
\label{sec:inter}

\begin{figure}
\includegraphics[width=80mm]{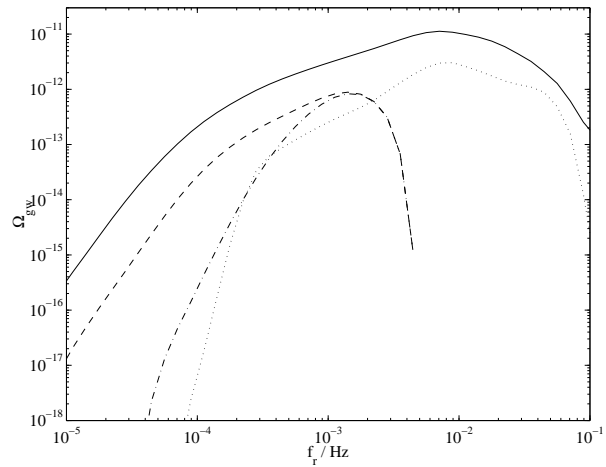}
\caption{The spectrum due to interacting binaries, for our fiducial Model
  A. The solid line shows the total WD--WD binary contribution, while
  the dotted line gives the spectrum from interacting binary WD--WD
  systems. The dashed line is the total nHe--WD spectrum, of which
  the dash-dot line gives the interacting nHe--WD contribution.}
\label{fig:inter}
\end{figure}

\begin{figure}
\includegraphics[width=80mm]{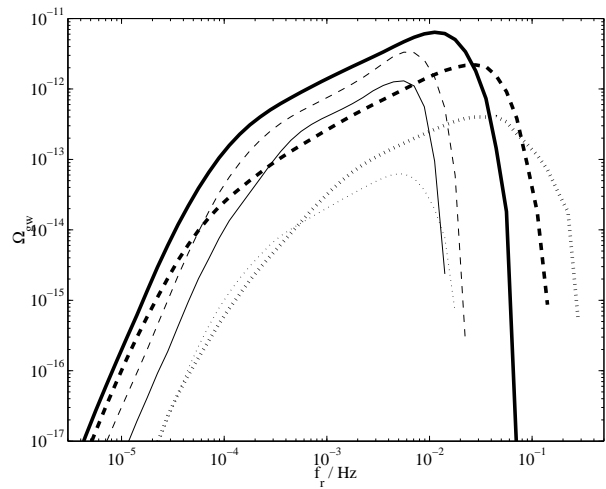}
\caption{The background received from different WD--WD pairings, for
  Model A. From top to bottom at 1 mHz: thick solid line: CO--CO, thin
  dashed line: He--CO, thin solid line: He--He, thick dashed line:
  CO--ONe, thick dotted line: ONe--ONe, thin dotted line: He--ONe.}
\label{fig:wdtypes}
\end{figure}

Interacting binaries (those in which either a WD or nondegenerate naked helium (nHe) star is
transferring mass on to a WD) contribute 13 per cent of the GW background at
1 mHz in Model A. Since at this frequency the majority of nHe star
companions fill their Roche lobes, most of the nHe--WD background
comes from interacting systems. At 10 mHz, 26 per cent of the GW signal comes
from interacting binaries, all of these necessarily WD-donor
systems. The GW spectrum due to interacting binaries is compared with
the total signal in Fig. \ref{fig:inter}.

The percentage contribution from interacting systems is
fairly constant across models, except for Model H, in which an
accreting CO WD is permitted to accumulate 0.3 M$_\odot$ of helium before
detonation, as opposed to the 0.15 M$_\odot$ in our fiducial
model. This increase in survival rate boosts the interacting binary
signal at 1 mHz by a factor of two. For the other models, the
interacting WD--WD signal is boosted when the WD--WD pairs formed
typically have larger mass ratios, so that more systems can commence
stable transfer upon Roche contact, e.g. Model C.

The spectral shapes from interacting systems are governed by the
mass-radius relation of the Roche lobe-filling star, and so do not
share the spectral slopes displayed by the detached binaries. The overall
contribution from interacting pairs is sufficiently small, however,
that the total spectral shape is little affected by their
presence. This is in line with results for the Galaxy found by
\citet{hb00} and \citet{nel01c}.

We can predict the spectral shape due to interacting WD--WD binaries
using some simple scaling relations (in the notation of section
\ref{sec:anal}): for a Roche lobe-filling WD of mass $M_{\rm{d}}$, we
have $M_{\rm{d}}^{-1/3} \propto R_{\rm{d}} = R_L \propto a
M_{\rm{d}}^{1/3} \propto M_{\rm{d}}^{1/3} f^{-2/3}$, using Kepler's
law (for conservative mass transfer). If we then assume that the mass of the donor WD is much less than
that of the accretor, then the system chirp mass ${\cal M} \propto  M_{\rm{d}}^{3/5}$, so
that $f \propto {\cal M}^{3/5}$ and the system gravitational wave
luminosity $L_{\rm{gw}} \propto f^{10/3} {\cal M}^{10/3} \propto
f^{16/3}$.

For sources sweeping (backwards) through frequency space, we have $N(f) \propto
1/\dot f \propto {\cal M}^{-5/3} f^{-11/3} \propto f^{-14/3}$.

Putting these together, we then have, for the emitted flux in the
logarithmic frequency interval around $f$, $\Omega_{\rm{gw}}(f)
\propto f F(f) = f L_{\rm{gw}} N(f) \propto f^{5/3}$. From
Fig. \ref{fig:inter}, we measure the spectral slope between 0.4 and 6
mHz to be $\sim$1.7, in good agreement with this
calculation. Interacting WD--WD sources are not present below this
frequency range because evolution to these frequencies requires more
than a Hubble time. Above $\sim 6$ mHz, the spectral shape depends on the
fraction of sources of high enough mass to radiate at a given
frequency; this number drops rapidly with increasing frequency. Note
that, within the 0.4--6 mHz range, since the spectrum $\Omega_{\rm{gw}}
\propto f^{5/3}$ for interacting WD--WD binaries rises relative to $\Omega_{\rm{gw}} \propto
f^{2/3}$ for inspiralling detached binaries, interacting
binaries are more important contributors at high frequencies than at low.

\subsubsection{WD types, chirp mass and merger rates}

The dominant component of the background at frequencies 0.1--10 mHz
comes from the inspiral of WD--WD systems. From Table
\ref{tab:wdtypes}, we see that approximately half of this
background comes from CO--CO pairs, descended primarily from higher
mass progenitors than the majority of He--He systems. The dominance of these systems is a result
of both the shorter time delay between star formation and DD birth for
more massive MS stars, and the larger chirp masses for CO--CO systems,
since the flux in the inspiral part of the
spectrum scales as $\Omega_{\rm{gw}} \propto f_{\rm{r}}^{2/3} {\cal M}^{5/3}$
(see section \ref{sec:anal}). These two factors outweigh the
fact that, from the IMF, many more potential progenitors of He WDs are
born than those which always produce CO or ONe WDs after envelope loss.

\begin{figure}
\includegraphics[width=80mm]{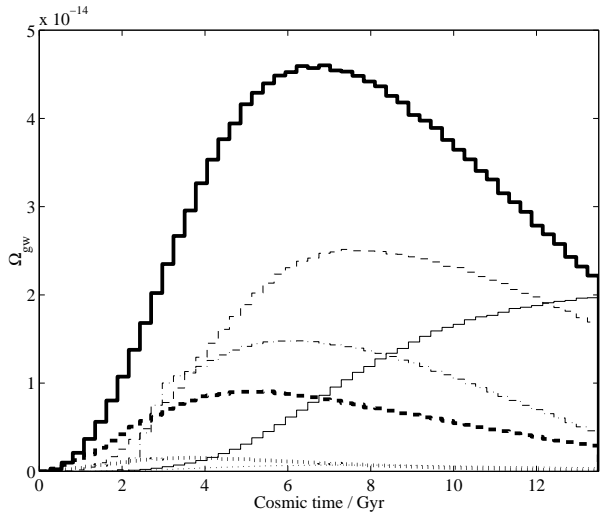}
\caption{The contribution to $\Omega_{\rm{gw}}$(1 mHz) received today
  as emitted from each shell of cosmic time, $\Delta T = T_0/50$, from
  each source type, for Model A. Linestyles are as in Fig. \ref{fig:wdtypes}, with the
  addition of the thin dash-dot line for nHe--WD pairs.}
\label{fig:wdtypes_time}
\end{figure}

Figure \ref{fig:wdtypes_time} shows however that, as more low-mass MS
stars evolve to the DD stage, the relative contribution to the GW
luminosity density from pairs involving He WDs is rising, and will
eventually dominate. The percentage contribution to the local ($z=0$) WD--WD
GW emission at 1 mHz from pairs including at least one He WD is 55 per cent, whereas
their contribution to the integrated cosmological background received today is
only 36 per cent.

\begin{figure}
\includegraphics[width=80mm]{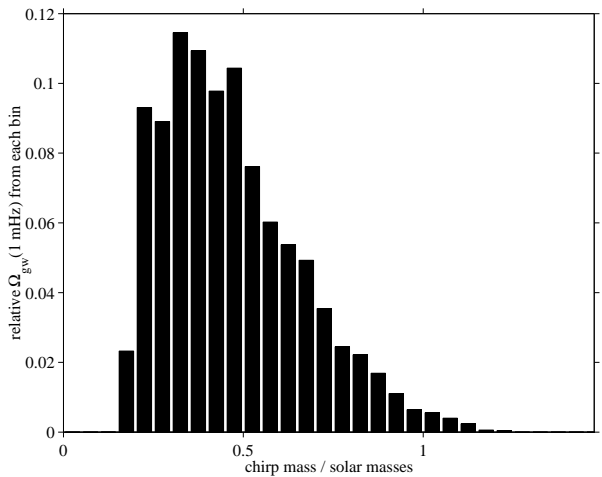}
\caption{Relative contribution to $\Omega_{\rm{gw}}$ at 1
  mHz as a function of chirp mass, for Model A, giving a mean
  flux-weighted chirp mass $\left< {\cal M} \right> = 0.45 \rm{~
  M}_\odot$.}
\label{fig:chirp_distn}
\end{figure}

\begin{figure}
\includegraphics[width=80mm]{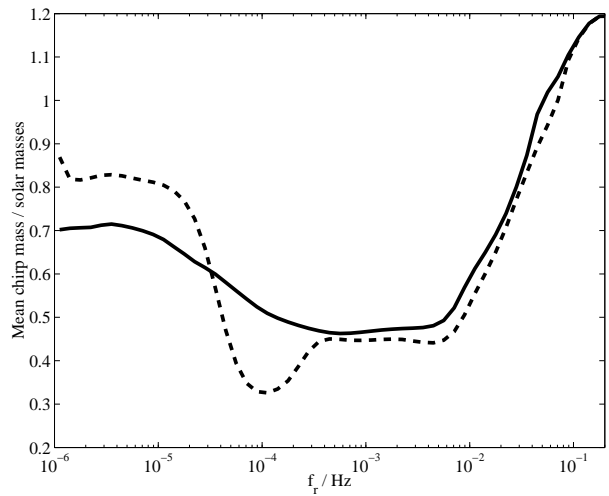}
\caption{Flux-weighted chirp mass contributing to the GW
  background received, as a function of frequency, for Model A. Solid
  line: detached WD--WD pairs only, dashed line: all source types. The
  dip seen in this curve around 0.1 mHz is due to low-mass main
  sequence stars.}
\label{fig:chirp_f}
\end{figure}

\begin{figure}
\includegraphics[width=80mm]{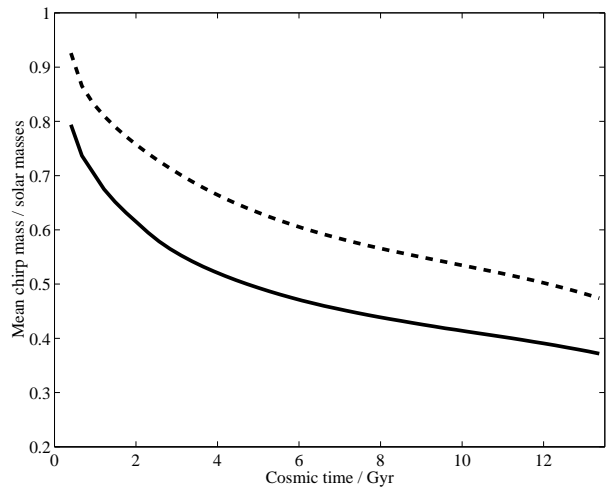}
\caption{The flux-weighted mean chirp mass contributing to emission received today at 1--3 mHz (solid line) and 3--10 mHz
  (dashed line), from each shell of cosmic time. All for Model A.}
\label{fig:mc_time}
\end{figure}

A useful way to look at this is through the chirp mass
distribution. Shown in Fig. \ref{fig:chirp_distn} is the
contribution to $\Omega_{\rm{gw}}$ at 1 mHz as a function of system chirp mass (defined in
section \ref{sec:anal}) for Model A, giving a flux-weighted mean chirp mass
of $0.45 \rm{~M}_\odot$.  As increasingly lower mass systems evolve off
the main sequence and become close DD pairs, this mean chirp mass is
decreasing with time, as shown in Fig. \ref{fig:mc_time}. The chirp
mass distribution depends on GW frequency (Fig. \ref{fig:chirp_f}), most notably shifting
towards higher masses at frequencies above which lower-mass WD--WD
pairs will have merged. The mean chirp mass is somewhat higher below the
critical spiral-in frequency, since for $f_{\rm{e}} < 2 \nu_{\rm{crit}}$, we
have $\Omega_{\rm{gw}} \propto {\cal M}^{10/3}$, and above
$2 \nu_{\rm{crit}}$, $\Omega_{\rm{gw}} \propto {\cal M}^{5/3}$ (see
section \ref{sec:anal}).

\citet{phi02} derived a simple expression for the GW background in terms
of the chirp mass ${\cal M}$, assumed constant across all sources, and the current space
density $N_0$ of remnant spiralled-in sources (with a weak dependence on cosmology
and star formation history). We can assess the usefulness of this
formula as a predictor of the background flux by using the
results of our population synthesis calculations to see whether the
computed fluxes can indeed be described by these two parameters only.

To calculate the
remnant density $N_0$, we first calculate the source spiral-in rate as a function of cosmic time. The rate of occurrence of Roche lobe contact between WD--WD pairs (we shall call
this the spiral-in rate)
is different from the rate of WD--WD mergers, since for some subset
of systems (those with mass ratios $q < 0.628$) stable mass transfer
will commence upon overflow, and an AM CVn system
will form. We keep track of both of these rates here.

For $f$ greater than both $2 \nu_{\rm{max}}$ and $2 \nu_{\rm{crit}}$, i.e. in the part of the
spiral-in regime above which sources are born (see section \ref{sec:anal}),
then for a quasi-constant spiral-in rate $\dot N$ over the timestep $T_0/50$, the continuity equation
(Eq. \ref{eq:cont}) simplifies to
\begin{equation}
\dot N = \sum_i \dot \nu N_i,
\end{equation}
summed over all sources $i$ at any given frequency satisfying the
above requirement. For each source, $\dot \nu$ is given
 by Eq. \ref{eq:fdot}. We perform this sum at each step in cosmic
 time, using systems with orbital
frequencies in the range $0.8 < \nu < 1.6$ mHz, which is above the maximum
injection frequency for the majority of sources, and below those
frequencies at which the lowest mass WDs are coming into contact. We
note that the inspiral time from $\nu \sim $ 0.5 mHz is less than $T_0/50=0.27$
Gyr for all ${\cal M} \ga 0.05 \rm{~M}_\odot$, so that at each timestep
 we are accurately representing the spiral-in rate at that time. The
 only exceptions are very low chirp mass systems, which we neglect here anyway, since these will be
 interacting binaries, which are spiralling \emph{out}. We also neglect all nHe--WD pairs, since the evolution of
these systems is not governed exclusively by gravitational radiation,
but also via radial evolution of the nHe star, and also because Roche
lobe contact occurs for these systems within our frequency range.

\begin{figure}
\includegraphics[width=80mm]{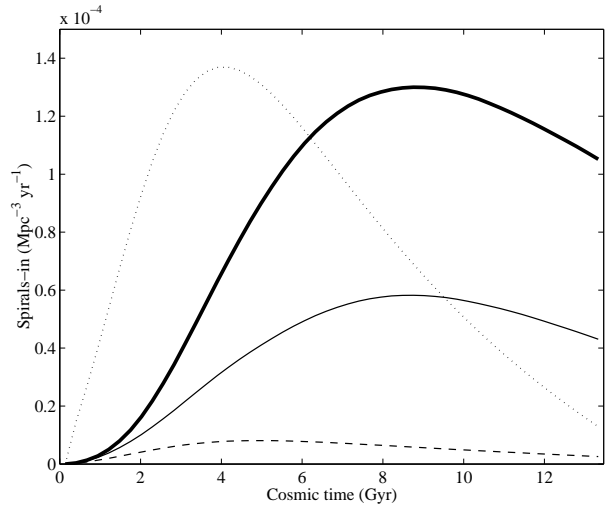}
\caption{The rate of WD--WD spiral-in as a function of cosmic
  time. The thick solid line gives the total spiral-in rate, while the
  thin solid line shows the merger rate, that is the inspiralling
  sources that will merge, and not commence stable mass transfer
  (i.e. become AM CVn binaries) upon
  Roche lobe overflow. The thin dashed line gives the rate of merger
  of WD--WD pairs with combined mass $> 1.4$ M$_\odot$. For reference,
  the cosmic star formation rate, multiplied by $1/(1000 \rm{M}_\odot)$,
  is plotted as the thin dotted line. All for Model A.}
\label{fig:mergers}
\end{figure}

The spiral-in and merger rates obtained from Model A are plotted in Fig.
\ref{fig:mergers}. The present-day remnant density $N_0$ needed for the formula of
\citet{phi02} is the time integral of the spiral-in rate, since this
gives the total number of sources that have contributed to the
background. From our calculated rate, we obtain
$N_0 = 1.17 \times 10^6 \rm{~Mpc}^{-3}$.

\citet{phi02} deals only with the GW emission from non-interacting
WD--WD systems, and so we should compare its predictions with
only the non-interacting component of our computed signals, in
addition to using a characteristic chirp mass ${\cal M}'$ for just those systems.
For Model A, our flux-weighted mean chirp mass for detached WD--WD
pairs is $\left<{\cal M}'\right> = 0.47 \rm{~M}_\odot$ at 1 mHz.
Eq. 16 of \citet{phi02}, converting to $h_{100}=0.7$, and
omitting the $\left<(1+z)^{-1/3}\right>$ scaling factor in the interests of
simplicity, becomes
\begin{equation}
\Omega_{\rm{gw}} = 1.1 \times 10^{-17} \left(\frac{{\cal
    M}'}{\rm{M}_\odot}\right)^{5/3} \left(\frac{N_0}{\rm{Mpc}^{-3}}\right) \left(\frac{f_r}{1 \rm{~mHz}}\right)^{2/3}.
\label{eq:phi}
\end{equation}
Using  $\left<{\cal M}'\right>$ and $N_0$ for Model A in the above, we
find $\Omega_{\rm{gw}}\rm{(1~mHz)} = 3.7 \times 10^{-12}$. We compare
this with the computed value for detached WD--WD pairs,
$\Omega_{\rm{gw}}(\rm{1~mHz}) = 3.0 \times 10^{-12}$, and note that these agree to
within 25 per cent. If we perform this same 
calculation for the other Models, we find that Eq. \ref{eq:phi}
overestimates the computed background by a similar fraction.

The \emph{variation} between models is thus well fitted by the
formula. The relative fluxes are reproduced by Eq. \ref{eq:phi} to within 5 per cent
for all Models except D and E, whose fluxes relative to Model A are overestimated by 7 and 16
per cent respectively. The dominant scaling is due to variations in $N_0$, since in most cases
$\left< {\cal M}' \right>$ varies little between Models. For the cases
in which $\left< {\cal M}' \right>$ does significantly change (D, E,
F, K and P), the omission of the
chirp mass scaling in Eq. \ref{eq:phi} can improve (D, E) or worsen
(F, K, P)
the agreement with the results of our detailed calculations. This is
perhaps as expected, since our flux-weighted chirp mass is in
fact not the same average as that required in the generalisation of
\citet{phi02} to accommodate a range of chirp masses. Such a value
would also incorporate the redshift-scaling omitted in the
above. We note, however, that neither $N_0$ nor either definition of ${\cal M}'$ is a
directly observable quantity, requiring as they do integrations over
cosmic time, and so are not easily determined from observations.

The computed spectral shape is not precisely $\Omega_{\rm{gw}} \propto
f_{\rm{r}}^{2/3}$ (see Fig. \ref{fig:basic}), so we do not expect an exact reproduction of the
spectrum using this formula. However, we conclude
that with a knowledge of $N_0$ and ${\cal M}'$, we can quickly
predict the detached WD--WD background amplitude and to some extent
its variation if these values change. We note however that a full population synthesis calculation enables the
inclusion of interacting systems, as well as the extraction of
detailed spectral shapes and source property distributions, which are
not available in a quick `manual' calculation.

\subsection{Progenitors}

Here we outline the relative contributions from the two main pathways
to the DD stage, and we assess the impact upon each of these routes of
varying the population synthesis model.

\begin{figure}
\includegraphics[width=80mm]{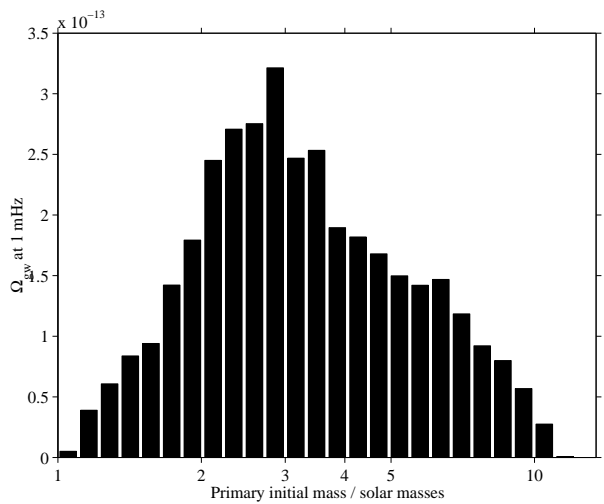}
\caption{Contribution to $\Omega_{\rm{gw}}$(1 mHz) received, as a function of
  ZAMS mass of the primary, for our fiducial Model A.}
\label{fig:m1init}
\end{figure}

Figure \ref{fig:m1init} shows the contribution to $\Omega_{\rm{gw}}$
at 1 mHz as a function of the initial mass of the primary, for Model A. The
descendants of primaries with ZAMS masses in the range 2--4 M$_{\odot}$ contribute 50 per cent of the
signal, the flux-weighted mean progenitor primary mass being 3.7 M$_\odot$. Most of the sources
in this range are the progenitors of CO WDs, since for $M \ga 2
\rm{~M}_\odot$, a CO WD will be
produced via a helium star upon envelope loss on the RGB, and
a CO WD will be produced directly if the envelope is lost on the
AGB. At 10 mHz, the mean progenitor mass rises to 4.7 M$_\odot$,
since the (necessarily more massive) WD--WD pairs contributing there are descended from only
the more massive ZAMS systems. The equivalent secondary mass distribution is not plotted here, but is always peaked towards initial mass ratios of unity.

\begin{figure}
\includegraphics[width=80mm]{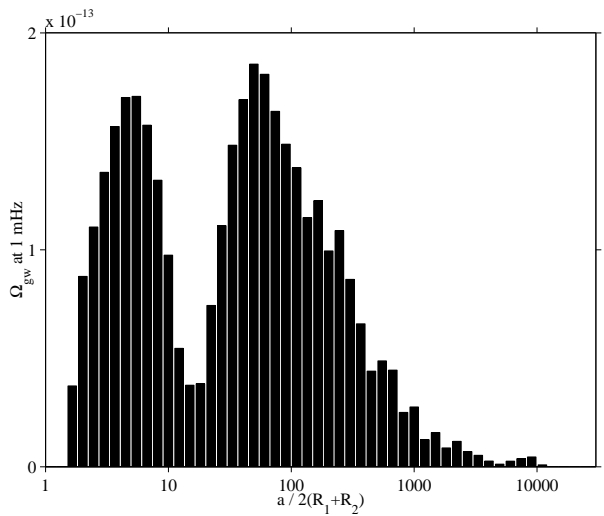}
\caption{Contribution to $\Omega_{\rm{gw}}$(1 mHz) received, as a function of
  initial progenitor semimajor axis, expressed as a ratio of the initial semimajor axis to the
  minimum separation permitted in the code. For Model A.}
\label{fig:ainit}
\end{figure}

Of perhaps more interest is the distribution in initial orbital
semimajor axis (Fig. \ref{fig:ainit}, for Model A), which has a clear bimodal
form, the peak at $a \sim 5 a_{\rm{min}}$ corresponding to DDs which
formed via RLOF+CEE, and the peak at $a \sim 50 a_{\rm{min}}$
corresponding to the CEE+CEE route. We can
therefore approximately determine the relative contributions from
these two routes by dividing this distribution between the two peaks
(at $a \sim 10 a_{\rm{min}}$ for most models); the result of this division for each model is shown
in Table \ref{tab:summary}. We note that the location of the CEE+CEE peak at
$a \sim 50 a_{\rm{min}}$ and the typical masses of the dominant
progenitor stars mean that for this route, the dominant
pathway involves primary overflow on the AGB, followed by secondary
overflow on its RGB.

For Model A, $\Omega_{\rm{gw}}\rm(1~mHz) = 1.4 \times 10^{-12}$ ($\sim
38$ per cent of the total) comes from sources that evolved via the RLOF+CEE
pathway. Since the WD--WD pairs from this route are generally more
massive than CEE+CEE pairs, the percentage contribution at 10 mHz from
this route rises to 44 per cent.

In general, we shall find that it is the RLOF+CEE contribution that is affected
more by varying the population synthesis model. Although it can be
affected significantly by varying the \emph{form} of the common
envelope prescription (Models E and P), the CEE+CEE signal is
quite robust to changes in the common envelope efficiency, since
if systems originating at one separation happen to coalesce in a
common envelope phase, using a given model, there exists a shell of
sources at greater $a$ to take their place as the closest WD--WD
systems at birth, out to a maximum of $a \sim 10^3 a_{\rm{min}}$ at
which Roche lobe overflow no longer occurs on the RGB or AGB. \citet{web98}
describe this effect in terms of shifting the `window' in initial
parameter space from which the closest DD systems are descended. The weak
dependence of results upon the common envelope efficiency parameter is
also seen in population synthesis calculations for other types of
binary, e.g. \citet{kal98} for LMXBs.

Returning to the DD case, the RLOF+CEE
pathway has no similar resource, occurring only in the rather
narrow range of initial separation in which RLOF commences in the
Hertzsprung gap. If we destroy more of these sources in the
ensuing CEE phase, we lose more of the contributions from the
RLOF+CEE route. 

Decreasing $\alpha$ (Model C) has this kind of deleterious effect upon the
RLOF+CEE pathway, but slightly increases the signal from CEE+CEE sources, since
the systems which survive to the close DD stage were on average more
widely spaced than for $\alpha=3.0$, so that the giant stars were
physically larger, i.e. more evolved, on average upon Roche lobe overflow, so gave
rise to more massive WDs (also with more widely differing masses). This corresponds to moving the second peak
in Fig. \ref{fig:ainit} to higher $a$. The lower mean chirp mass is
largely attributable to the increased number of low-chirp mass interacting binaries
present at this frequency, since the typical WD--WD mass ratio is
larger, as described in section \ref{sec:inter}.
 Increasing the efficiency
parameter (Model D) has the opposite effect upon each route. If on the other hand we
use the common envelope formalism of \citet{nel00} (Model E), it becomes less
simple to disentangle the two routes, since now they overlap somewhat in
initial $a-$space, but since we know that this modification ought not to
affect the RLOF+CEE contribution, we hold this fixed from Model A. The new CEE+CEE value turns out to be significantly
enhanced, since a wider range of initial separations has been opened
up to double common envelope survival. The nearer (by design) equality
of WD pair masses leads to a decrease in the number of WD--WD AM CVn
systems produced, and hence a smaller contribution from interacting
systems than for Model A.

The envelope ionisation energy becomes a significant part of the
energy balance in AGB stars, and so its inclusion is important in common envelope
phases that commence at large orbital separations. Increasing the
fraction of this energy included in the envelope binding energy (Model
N) therefore decreases the number of wide binaries able to shrink
enough to form close WD--WD pairs. Omitting it entirely (Model O),
thus increasing the envelope binding energy, has the opposite effect.

Model P shows the greatest departure from Model A in terms of GW flux and
mean chirp mass. The progenitor mass distribution for Model P is
peaked towards higher mass (6 -- 8 M$_\odot$) stars than for other
Models. These differences can be traced to the outcome of common
envelope phases on the Hertzsprung Gap (HG). The \textsc{bse} fitting
formula returns values of $\lambda$ substantially smaller than 0.5 for
most HG stars, corresponding to a high degree of central
concentration. Therefore using $\lambda=0.5$ in Model P results in
much less shrinkage in these situations.

High mass stars expand in radius by a large factor in
their HGs, so that the final Roche contact (for both pathways) is
often a common envelope phase involving a HG star. The survival rate from this CE
phase is boosted by the fixed lambda as described above. The resulting
GW flux is therefore also
greatly boosted for these higher mass stars, whose descendent WDs are
sufficiently massive that relatively few are required to dominate the
background GW flux. Given however that small
values of lambda are robust for HG stars (they are also seen in the
calculations of \citealt{dew00}), we choose not to consider
this prescription as a reasonable uncertainty on the background.

The lower chirp mass $\left<{\cal M}\right>$ seen for Model H is due to the inclusion of an
increased number of interacting sources at 1 mHz, compared with Model
A; the value $\left<{\cal M}'\right>$ appropriate for just detached WD--WD pairs
for this model (used in the previous section) is the same as for Model A.

Starting all systems with circular orbits (Model K) boosts the
RLOF+CEE pathway, because fewer systems given initially tight orbits
are lost due to immediate collision at periastron. Since systems
descended from the RLOF+CEE route are generally higher-mass, the mean
chirp mass for Model K is higher than for Model A. CEE+CEE route
systems are little affected; the high-$a$ peak in Fig.
\ref{fig:ainit} is simply narrowed in $a$-space, since orbital separations
are no longer altered by tidal circularisation before Roche
contact.

Aside from orbital circularisation, the main role of tides in the
evolution to the DD stage is in orbital shrinkage before Roche
contact, due to spin-up of the giant star. Neglecting tidal effects is
thus similar to increasing the common envelope
efficiency parameter, i.e. the progenitors of close DDs from the
CEE+CEE route have smaller
initial orbital separations, and the DDs produced have smaller chirp
masses on average. If on the other hand tidal effects were much
stronger than in the \textsc{bse} code, then we would expect little
impact upon this route, since giant-star corotation is already
typically achieved before Roche lobe overflow with the tides in \textsc{bse}.

The CEE+CEE route is as expected largely unaffected when we perturb
dynamically stable mass transfer on the Hertzsprung gap. Much as one might expect,
the RLOF+CEE route is enhanced when one enhances the transfer on the
Hertzsprung gap (Model L), so that more mass is transferred to the
companion, and more systems avoid a common envelope phase during the
first phase of mass transfer (which tends to lead to merger). The
orbit is also widened to a greater extent during transfer, meaning
that more systems will survive the common envelope phase when the
secondary evolves. Making the transfer semi-conservative (Model M) has
an opposing effect; the orbit is widened less during stable overflow,
meaning that more systems are destroyed in the ensuing common envelope phase.

The steeper Scalo IMF (Model F), normalised to the local space
density of low-mass stars, produces fewer intermediate (and high) mass
stars than the KTG IMF, and so fewer of the dominant progenitors in Fig.
\ref{fig:m1init} are produced. More of the compact binaries are then
descended from lower-mass progenitors than for Model A, giving rise to
their lower mean chirp mass.
If we had instead normalised to the local core-collapse supernova
rate, as in \citet{sch01}, we would instead have ended up with a
correspondingly \emph{higher} background from Model F.

Altering the shape of the cosmic star formation history (Model J) has
little impact upon the background, since most of the sources
contributing have MS evolution times of less than a few Gyr (see
Fig. \ref{fig:m1init}). This is a
strong argument in favour of using an integral constraint (such as IR
luminosity density), and not a present-day constraint (such as
local core-collapse supernova rate), since normalising according to
the supernova rate introduces a strong dependence on the shape of the
cosmic star formation history curve, through the difference in
amplitude between the local rate and the rate at the peak of star
formation, which can easily skew the overall normalisation.

Finally, Models Q and R lead to larger gravitational wave backgrounds
than Model A, mainly because lower metallicity stars tend to leave the
main sequence earlier, and thus a greater fraction of the stellar mass
in the universe today is present in the form of remnants. The
difference in received flux is, however, slight, on the order of 10
per cent. We conclude that keeping detailed track of abundance
variations is not essential to calculation at the present level of accuracy.

\section{Outlook}
\label{sec:opt_pes}

\begin{table}
\caption{Summary of the properties of the optimistic and pessimistic
  models, along with the fiducial Model A}
\begin{tabular}{cccccc}
\hline
Model & \% DD & $\left<q\right>$ & $\Omega_{\rm{gw}}\rm{(1~mHz)}$ &
$\left<{\cal M}\right>$ & $N_0$\\
\hline

Optimistic & 26 & 0.75 & 5.99 & 0.46 & 1.85 \\
A & 18 & 0.62 & 3.57 & 0.45 & 1.17 \\
Pessimistic & 14 & 0.66 & 0.95 & 0.40 & 0.32 \\
\hline
\end{tabular}
\label{tab:opt_pes}
\end{table}

Based on the above indications of which effects boost the GW background and
which reduces it, we construct two models in an attempt to put
upper and lower limits on the background we predict. Our use
of the terms `optimistic' and `pessimistic' assumes that this
background constitutes signal for the reader; if it constitutes a
noise, the nomenclature should be reversed.

\textbf{Optimistic model:}
This has the properties of Model A, except for: the \citet{nel00} common
envelope formalism, initially circular orbits, enhanced mass
transfer on the HG, edge lit detonations only after accretion of 0.3
M$_\odot$ and no ionisation energy in envelope binding energies used for
common envelope phases. Note that some of these individually boosting
effects do not make a double-boost in combination; for example the no spiral-in
common envelope prescription tends to lead to DD mass ratios closer to unity,
which means that fewer systems undergo stable mass transfer upon
contact, and so the enhancement brought by the higher ELD limit is
less effective in increasing the amplitude of the background. We also
include the estimated error on our overall normalisation (see section
\ref{sec:csfh}), by using a cosmic star formation rate everywhere 30 per cent higher than our
fiducial one.

\textbf{Pessimistic model:}
The pessimistic model contains the elements found in the previous
section to decrease the amplitude of the GW background. The properties of this model
are thus the same as Model A, except for:  $\alpha=2.0$, Roche lobe
overflow is semiconservative on the HG, the Scalo initial
mass function is used and 100 per cent of the ionisation energy is
included in the envelope binding energies used in common envelope phases. In addition, we use a star formation rate
everywhere 30 per cent lower than
our fiducial one, in our cosmological integral.

\begin{figure}
\includegraphics[width=80mm]{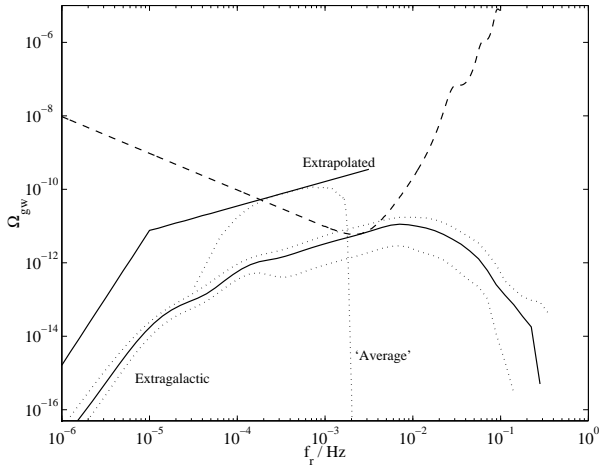}
\caption{Optimistic (upper dotted), fiducial (Model A, lower solid line) and pessimistic
  (lower dotted) extragalactic backgrounds plotted
  against the LISA (dashed) single-arm Michelson combination sensitivity curve (see
  http://www.srl.caltech.edu/$\sim$shane/sensitivity/). The `unresolved' Galactic close
  WD--WD spectrum from \citet{nel01c} is plotted (with signals from
  binaries resolved by LISA removed), as well as an
  extrapolated total, in which resolved binaries are
  restored, as well as an approximation to the Galactic MS--MS
  signal at low frequencies.}
\label{fig:lisa}
\end{figure}

These prescriptions were used to create Galactic DD populations, which
were found to compare reasonably with observations. Then the cosmological
integrals were carried out for each. The results of this are
summarised in Table \ref{tab:opt_pes}, and the optimistic, fiducial and
pessimistic total background spectra are plotted in Fig. \ref{fig:lisa}
along with the LISA sensitivity curve, and the Galactic WD--WD
background taken from \citet{nel01c}.  We plot both the `unresolved' (`average')
background curve from their paper, which is for DD pairs only, with the
resolved sources removed, and an extrapolated `total'
background. In this we have added back in the resolved close binaries
and made an approximation to the MS--MS contribution at lower
frequencies, in an attempt to represent the Galactic signal over the
full frequency range plotted.

\begin{figure}
\includegraphics[width=80mm]{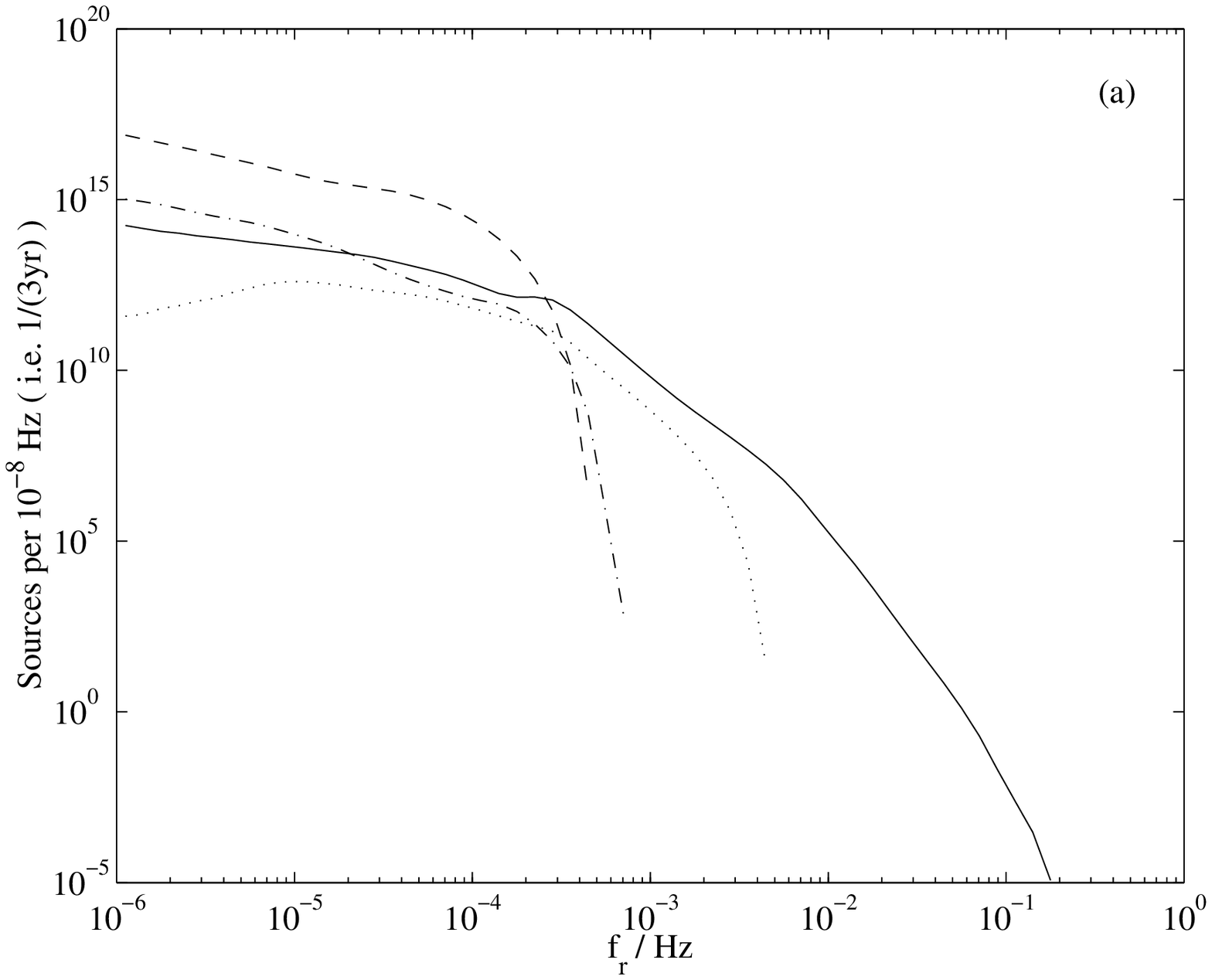}
\includegraphics[width=80mm]{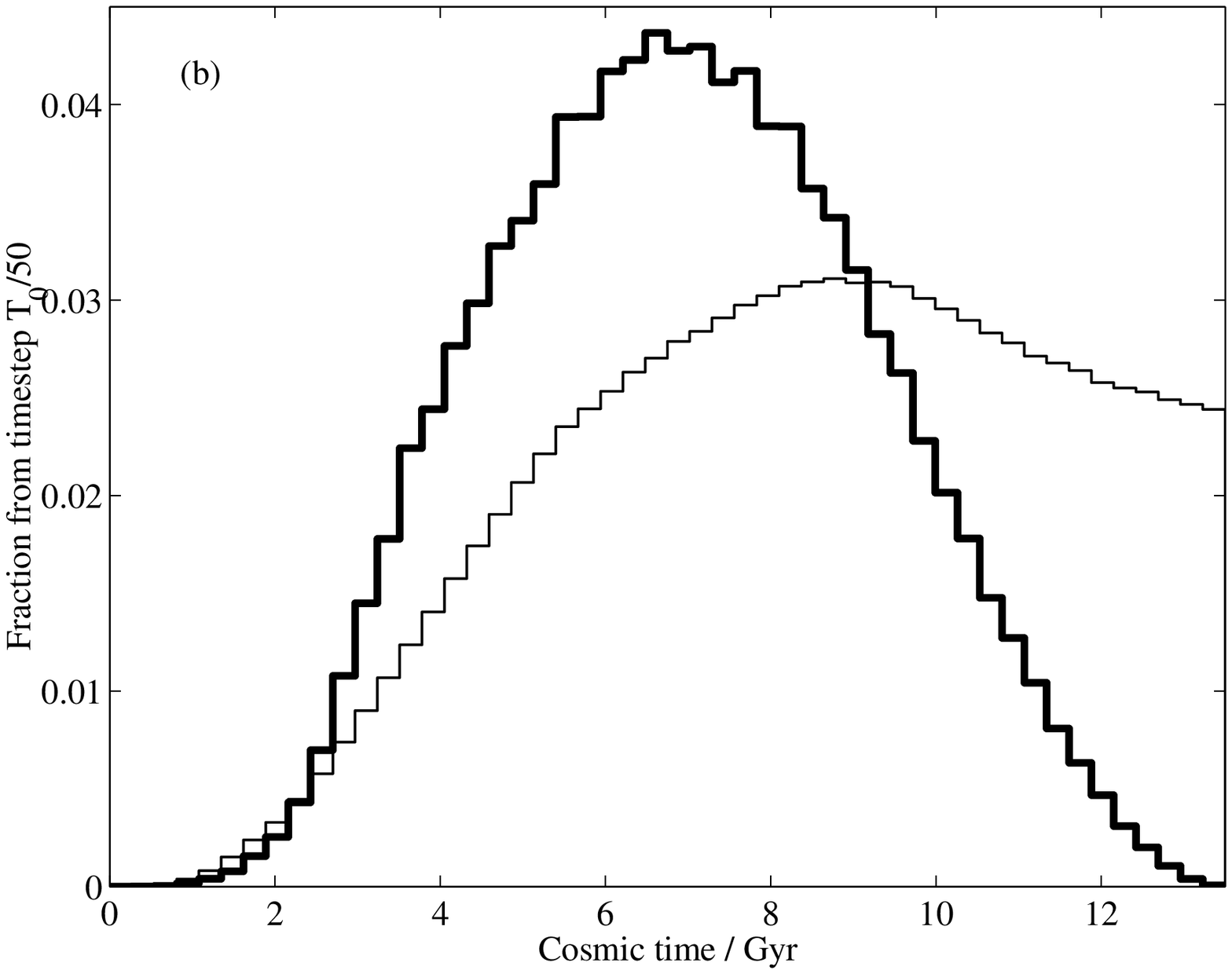}
\caption{(a) The number of systems per $10^{-8}$ Hz contributing to the cosmological
GW background as received today. Linestyles denote the evolutionary
classes as in Fig. \ref{fig:basic}. (b) Thin line: the fractional
contribution at 10 mHz to the GW background as a function of cosmic
time (from shells of width $T_0/50$). Thick line: the same, but in
terms of the number of sources contributing to the flux received from
each cosmological time-shell.}
\label{fig:nsources}
\end{figure}

Plotted in Fig. \ref{fig:nsources}(a) is the number of systems per 1/(3
yr) frequency resolution element
contributing to the GW background as received
today. We see from this that at frequencies $f_{\rm{r}} \la 50$ mHz,
there will be too many individual WD--WD sources contributing in each
resolution element for this background to be completely resolved and
subtracted source by source by missions with plausible lifetimes. However, from Fig. \ref{fig:nsources}(b),
we see that much of the flux comes from relatively nearby sources, and
the WD--WD numbers drop rapidly above 50 mHz (leaving the lower
background from rare neutron stars and black holes, not considered in
this paper). Thus it may be possible for future missions more
sensitive than LISA to subtract this
background at high frequencies.

\section{Comparison with previous work}
\label{sec:prev}

\citet{hil90} and \citet{kos98} each made an order of magnitude estimate of the ratio of
the extragalactic to Galactic GW flux from DDs. In order to
facilitate comparison, and to compare like with like as far as
possible, we divide our calculated extragalactic flux at
1 mHz by the most recently calculated value \citep{nel01c} for the
Galactic flux at the same frequency (which is a factor $\sim 3$
smaller than that found by \citealt{web98}). The correct curve from Fig.
\ref{fig:lisa} to use for this comparison is our `extrapolated'
curve. We find $\Omega_{\rm{extragal}}/\Omega_{\rm{gal}} =$
2.0 per cent at 1 mHz for Model A, with a range of 0.5 -- 3.4 per cent between
optimistic and pessimistic Models. 

\citet{hil90} predicted a factor $\sim$ 1.6 per cent (for an Einstein-de Sitter
universe with no cosmological evolution of galactic GW
luminosity). This estimate is in good agreement with our value.

\citet{kos98}, on the other hand, predicted that, for a cosmology of the
type used in this paper, the extragalactic background should be of order 10 per cent of
the Galactic one, when one takes into account the evolution of star
formation rate with redshift. This result is in clear disagreement with our
findings, but this can be explained by noting that their ratio is
artificially raised by a number of factors: first, the fiducial
scalings of $\Omega_{\rm{b}}$, $\left< r \right>$ and $h_{100}$
in their eq. 13 are higher than their true values, boosting the extragalactic signal. Second, the same
star formation rate as a function of redshift was used for different
cosmologies, which leads to an
artificial boost to the lambda-cosmology extragalactic flux \citep*[see e.g.][]{som01}. Lastly, the cosmic
star formation rate adopted was not normalised to any integral
constraint, but merely to the current star formation rate. All of
these factors lead to their calculation yielding a misleadingly high
extragalactic contribution to the GW background.

\citet{sch01} made a more direct calculation of the background. At 1 mHz, their derived background
level (for $h_{100}=0.7$) is $\Omega_{\rm{gw}} = 1.2 \times 10^{-11}$, with
no quoted uncertainty on this value. This lies a factor two outside of
our predicted
range for the background. The discrepancy can be understood mainly in terms
of their different method of normalisation: they normalised to the
local core collapse supernova rate, and used the steep Scalo IMF,
meaning that more low- and intermediate-mass stars were born in their
simulations than measured by \citet{col01}. As explained in
Section \ref{sec:csfh}, we believe that normalising to an integral
constraint on the birth of low-mass stars is a more robust
method. \citet{sch01} also used a binary fraction of 100 per cent, cf. our 50 per cent.

The \emph{shape} of the spectrum in \citet{sch01}, however, we cannot
explain. The spiral-in part of the spectrum ($f_{\rm{r}} \ga 10^{-4}$ Hz) has
the form expected from Section \ref{sec:anal}, but the
static regime instead displays a prominent `bump' at frequencies
($f_{\rm{r}} \sim 3 \times 10^{-5}$ Hz) just
below the transition to the spiral-in regime, the amplitude of which decays rapidly towards
lower frequencies. No such feature is seen in our calculated spectra. This type of feature is difficult to explain in
terms of the arguments in Section \ref{sec:anal}, unless the vast majority of WD--WD pairs are
born precisely into this `bump', which seems unlikely, since the same feature is
seen for all types of compact object pair (e.g. NS--NS, NS--BH),
despite their very different formation routes.
\section{Conclusions}
\label{sec:conclusion}

We predict that the background of gravitational waves from extragalactic
binary stars is

\begin{enumerate}
\item Dominated by double main sequence binaries for $f_{\rm{r}}<10^{-4}$ Hz.
\item Dominated by double white dwarf binaries for $10^{-4}<f_{\rm{r}}<10^{-1}$Hz.
\end{enumerate}

Concentrating on the spectrum around $1$~mHz:

\begin{enumerate}
\item The fraction of critical density in gravitational waves received
  in the logarithmic frequency interval around 1 mHz lies in the range
  $1 \times 10^{-12} < \Omega_{\rm{gw}} < 6 \times 10^{-12}$, with the
  most likely value in the range $3-4 \times 10^{-12}$.
\item The flux-weighted mean chirp mass of the contributing binaries
 is $\langle{\cal M}\rangle = 0.45\rm{~M}_\odot$.
\item Half of the background comes from binaries whose more massive
 (primary) star had a mass in the range 2--4 $\rm{M}_\odot$ (and $\sim$70 per
 cent from primaries originally less massive than 4 M$_\odot$). The estimate
 of the background is thus more robust to uncertainties in the IMF and
 mass cuts if normalised to the present
 density of starlight than if normalised to core-collapse supernova rates.
\item $\sim 60$ per cent of the GW signal is from binaries with initial
 semi-major axes in the range of 30--1000 stellar diameters, in
 which the Roche contact of both primary and secondary stars led
 to unstable transfer and a common envelope. The background level produced
 by these systems is quite stable against uncertainties in the
 efficiency of the common envelope phase, though the signal can be
 changed somewhat through use of a non-standard common envelope
 prescription.
\item $\sim 40$ per cent of the GW flux comes from systems descended from 
 binaries with initial semi-major axes of about 5 stellar
 diameters, in which the first Roche contact occurred in the Hertzsprung
 gap, with stable overflow, but the second Roche contact led
 to unstable transfer and a common envelope. The background level
 produced by these systems is sensitive to uncertainties in
 common envelope and mass transfer physics.
\item interacting systems (AM CVn binaries) contribute only
 about 10 per cent of the energy density in gravitational waves.
\end{enumerate}

The above holds true for $0.5 \la f_{\rm{r}} (\rm{mHz}) \la 5$. Above this
range, as the lower-mass WD--WD pairs reach contact and drop out of
the spectrum due to mergers, the properties change (values at 10 mHz
in the parentheses which follow): the contribution
from interacting binaries increases (26 per cent), the RLOF+CEE
route contribution (44 per cent) and the mean primary progenitor mass increase
(4.7 M$_\odot$) and the mean chirp mass is higher (0.56 M$_\odot$).

We find that at all frequencies, our derived spectral shape can be
understood in terms of simple arguments, and that this shape is
essentially independent of the population synthesis model used.

\section*{Acknowledgements}

This research has been supported in part by NASA grant NAG5-10707.
We thank J. Hurley for supplying the original BSE code,
and for his advice and insights, and we also thank G. Nelemans and
B. Hansen for useful discussions. We are also grateful to the referee,
V. Kalogera, for her valuable comments.

\bsp

\end{document}